\documentclass[%
 preprint,
 amsmath,amssymb,
 aps,
]{revtex4-2}

\usepackage{graphicx}% Include figure files
\usepackage{dcolumn}% Align table columns on decimal point
\usepackage{amsmath,bm}
\usepackage{epstopdf,epsfig}
\usepackage{newtxtext}
\usepackage{newtxmath}
\usepackage{natbib}
\graphicspath{{Figures/}}
\usepackage{hyperref}
\hypersetup{
	colorlinks = true,
	urlcolor   = blue,
	citecolor  = black,
}

\newcommand{\RomanNumeralCaps}[1]
 \linenumbers

\usepackage{subcaption}
\usepackage{placeins}

\newcommand{\hq}{\bar{\bf q}}

\newcommand{\stau}{\underline{\underline{\tau}}}
\newcommand{\p}{\partial}

\DeclareOldFontCommand{\rm}{\normalfont\rmfamily}{\mathrm}
\DeclareOldFontCommand{\sf}{\normalfont\sffamily}{\mathsf}
\DeclareOldFontCommand{\tt}{\normalfont\ttfamily}{\mathtt}
\DeclareOldFontCommand{\bf}{\normalfont\bfseries}{\mathbf}
\DeclareOldFontCommand{\it}{\normalfont\itshape}{\mathit}
\DeclareOldFontCommand{\sl}{\normalfont\slshape}{\@nomath\sl}
\DeclareOldFontCommand{\sc}{\normalfont\scshape}{\@nomath\sc}
\DeclareRobustCommand*\cal{\@fontswitch\relax\mathcal}
\DeclareRobustCommand*\mit{\@fontswitch\relax\mathnormal}
\DeclareMathAlphabet{\mathsfbi}{OT1}{\sfdefault}{bx}{sl}

\newcommand{\bn}{\bar{\eta}}

\newcommand{\Rint}{\mathcal{R}_\text{int}}
\newcommand{\dsint}
{\mathrm{d}s}% {\mathrm{d}s_{\mathrm{int}}}

\newcommand{\dAb}
{\mathrm{d} A_{\Omega_{xy}}}
\newcommand{\Rb}{\mathcal{R}^0_\text{int}}
\newcommand{\kmax}{k_{max}}
\newcommand{\smax}{\sigma^{max}_r}
\newcommand{\comsol}{COMSOL Multiphysics$^\text{TM}$}
\newcommand{\tens}[1]{\mathsfbi{#1}}
\begin{document}

\preprint{APS/123-QED}

\title{Stability of liquid film coating a horizontal cylinder: interplay of capillary and gravity forces
}% Force line breaks with \\

\author{Shahab Eghbali}
 \email{shahab.eghbali@epfl.ch}
\author{Simeon Djambov}%
\author{Fran\c cois Gallaire}
\affiliation{%
 Laboratory of Fluid Mechanics and Instabilities, \'Ecole Polytechnique F\'ed\'erale de Lausanne, Lausanne CH-1015, Switzerland
}%

\begin{abstract}
We study the drainage of a viscous liquid film coating the outside of a solid horizontal cylinder, where gravity acts vertically. We focus on the limit of large Ohnesorge numbers $Oh$, where inertia is negligible compared to viscous effects. We first study the evolution of the axially invariant draining flow, initiated at rest with uniform film thickness $\delta$. Non-linear simulations indicate that for each $\delta$, there is a threshold in the Bond number ($Bo$), which compares the gravitational effects with surface tension, above which the draining liquid bulk ruptures. This critical $Bo$ is found to scale inversely with
%the mean film thickness
$\delta$, defines the existence of a quasi-stationary pendant liquid curtain remaining sustained below the cylinder by surface tension. The interface of the pendant curtain is unconditionally linearly unstable and is prone to Rayleigh-Plateau-like, capillarity-driven, and Rayleigh-Taylor, gravity-driven, instabilities.
The linear stability of the quasi-static state along with an energy analysis of the unstable mode illustrates that while the Rayleigh-Taylor instability is always present, capillary effects dominate the instability at small $Bo$, which promotes the formation of pearls enveloping the cylinder. In contrast, at large $Bo$, capillarity acts in a stabilising way and the instability is purely gravity-driven, forming underside modulations.
We present the asymptotic energy repartition representing the different physical mechanisms at play in the instability of the saturated curtains for a wide range of $\{Bo,\delta\}$.
The results of the linear analysis agree with the pre-existing experiments of~\citeauthor{Bruyn1997} [Phys. Fluids \textbf{9(6)}, 1599 (1997)] and non-linear simulations of~\citeauthor{Weidner1997} [J. Colloid Interface Sci. \textbf{187}, 243 (1997)] in the limit of a thin film and extend the results for thick films.
Additionally, based on the volume made available for droplet growth by the development of the most linearly amplified wavelength, we build a tentative regime diagram that predicts the final patterns emerging from the pendant curtain, namely an array of saturated pearls or pendant drops or the onset of three-dimensional droplet pinch-off.
Furthermore, a transient growth analysis accounting for the time-evolution of the base state towards a saturated curtain conclusively demonstrates that the initial flow evolution does not result in altering the most amplified wavelength, thus rationalizing {\it a posteriori} the asymptotic analysis to predict the fate of the three-dimensional patterns.
\end{abstract}

\maketitle

\section{Introduction} \label{sec:intro}
Gravity-driven liquid films coating the outer side of a solid surface are ubiquitous in nature. Some daily-life examples are dew covering the strings of spider silk~\citep{Elettro2016}, and raindrop spread and accumulation around tree branches~\citep{Herwitz1987},
%and lava flow on volcanoes~\citep{Kilburn2000}, 
to name a few.
Such a flow is also of interest for several practical applications, for instance in coating industries~\citep{Quere1999,Shen2002,Duprat2007}, painting~\citep{Blair1969,Zenit2019}, vapor absorption~\citep{Chinju2000,Grunig2012,Hosseini2014,Ding2018flow}, desalination~\citep{Sadeghpour2019}, shell fabrication~\citep{Lee2016} and heat exchangers~\citep{Zeng2017,Zeng2018}. The motion of the coating liquid film may be influenced by several factors, most chiefly the solid surface geometry as well as the interplay of gravitational, capillary, and viscous forces. It is therefore not surprising that the dynamics of coating film flows, their instabilities, and subsequent pattern formation have gained attention in the last decades~\citep{Eggers2008,Gallaire2017}. 

%%%
A particular coating flow configuration is the gravity-driven flow around a horizontal solid cylinder. The interface of this draining film is prone to several instabilities which have been studied extensively, both numerically and experimentally.
A vast majority of the numerical analyses have focused on the thin film limit and exploited the lubrication approximation~\citep{Oron1997}.
\citet{Balestra2019} for instance investigated the conditions for fingering instability to occur during the spreading of a thin film over a partially wetting horizontal cylinder. Several other studies addressed the liquid motion, film thickness, and the centrifugal instability of a liquid-coated cylinder rotating around its axis~\citep{Hansen1994,Peterson2001,Ashmore2003,Evans2004,Li2018}. However,
%surprisingly, 
the coating flow around a fully wetting horizontal cylinder, in absence of rotation, has received little attention.

\citet{Reisfeld1992} investigated the two-dimensional isothermal and non-isothermal evolution of a thin liquid film of initially uniform thickness around a stationary horizontal cylinder.
Their study describes the flow as the drainage of the liquid film around the cylinder that leads to the formation of a pendant liquid curtain at the bottom of the cylinder. They also investigated the shape variation of the pendant interface as a function of the thermal properties of the flow.
\citet{Limat1992} investigated experimentally the gravity-driven dripping and jetting underneath a horizontal cylinder subject to a continuous flow feed
while
\citet{Bruyn1997} investigated experimentally the instability of a thin liquid layer that covers a horizontal cylinder and measured the early-time wavelength of the unstable patterns and observed that their growth was followed by possible droplet pinch-off at large times when surface tension is weak compared to gravity.
Using non-linear simulations of a similar isothermal flow configuration as in \citet{Reisfeld1992},~\citet{Weidner1997} demonstrated the three-dimensional temporal evolution of the flow. Starting from a core-annular state, initially perturbed by a low-level white noise, they depicted the flow evolution in four phases:
First, the axially invariant liquid bulk is pulled off without any evidence of instabilities. After a while past the establishment of a pendant curtain, the second phase starts with the appearance of growing small-amplitude longitudinal wavy perturbations at the bottom of the interface.
The third phase evidences non-linearities that affect the growth of these wavy patterns and lead to the formation of large drops separated axially by very thin liquid ridges, as well as possible smaller satellite drops.
In the fourth phase, the satellite drops may coalesce into bigger drops and the interaction between single drops ceases. \citet{Weidner1997} highlighted that in the limit of small cylinder diameter with respect to the capillary length, where surface tension is important, perturbations lead to an upward motion of the flow that results in pierced droplets surrounding the cylinder, hereafter {\it pearls}. Later,~\citet{Weidner2013} reported a similar reversal of drop formation due to surface tension modification in the presence of surfactants.
While these studies are limited to a thin film layer, where the lubrication approximation holds, the formation of a collected liquid bulk underneath the cylinder may violate the thin-film assumption.
\citet{Weidner1997} reasoned in analogy with a static pendant drop solution, where except for highly curved regions of the interface, a thin-film approximation results in a fair prediction of the curtain interface. 
% remarkably, at the south pole of the cylinder that is the thickest region, exhibits a good accordance.
%
Furthermore, despite their effort to apply an initial white noise to the film, the amplification of perturbations during the first phase (liquid pull-off), and its potential influence on the final droplet size are not well understood.
Characteristics of the emerging patterns and their shape strongly resemble those of the classical Rayleigh-Taylor instability~\citep{Rayleigh1882,Taylor1950} and capillary instability, which can be considered as an extension of the Rayleigh-Plateau instability~\citep{Plateau1873, Rayleigh1878} on a fibre~\citep{Quere1999}.

The present study revisits the flow configuration as in~\citet{Weidner1997} from an arbitrary thick-film viewpoint and aims at linking the pattern formation in such a flow with the linear interplay between the capillary-driven and gravity-driven instabilities.
This aim is pursued by means of the linear stability analysis of the quasi-static pendant liquid curtain for a wide range of parameters.
% Additionally, we examine the amplification of the initial perturbations during the liquid drainage around the cylinder by means of transient growth analysis and address its effects on the pattern formation.
%

This paper is structured as follows. The methodology is detailed in \S\ref{sec:horizontalfiber-Gov-eq}. The problem formulation and governing equations are presented in \S\ref{sec:horizontalfiber-prob-formulation}, from which the base flow is deduced and discussed in \S\ref{sec:horizontalfiber-baseflow}.
In \S\ref{sec:horizontalfiber-formulation-stability-analysis}, the formulation for the stability analysis and the linearised governing equations are elaborated. 
%\textbf{In \S\ref{sec:horizontalfiber-formulation-TG} the formulation of the transient growth analysis is detailed.}
The numerical methods are presented in \S\ref{sec:horizontalfiber_Numerical-method}. In \S\ref{sec:horizontalfiber_results}, the results of the stability analysis
%\textbf{and transient growth analyses} 
are presented.
In \S\ref{sec:horizontalfiber_results_LSA_overview} the influence of the Bond number on the characteristics of the flow stability is summarised.
In \S\ref{sec:horizontalfiber_res_energy}, the flow is investigated from an energy perspective, and the base flow and the perturbed flow's energy balance are detailed in \S\ref{sec:horizontalfiber_res_energy_Baseflow} and \S\ref{sec:horizontalfiber_res_energy_LSA}, respectively, followed by the obtention of an asymptotic energy diagram in \S\ref{sec:horizontalfiber_res_energydiagram}.
Finally, the linear stability regime diagram and the large-time pattern formation are discussed in \S\ref{sec:horizontalfiber_results_3dpinch}.
	
%%%%%%%%%%%%%%%%%%	
\section{Governing equations and methods} \label{sec:horizontalfiber-Gov-eq}
%%%%%%%%%%%%%%%%%%%%%%%%%%
\subsection{Problem formulation} \label{sec:horizontalfiber-prob-formulation}
%%%%%%%%%%%%%%%%%%%%%%%
\begin{figure}
	\centerline{\includegraphics[width=0.9\textwidth]{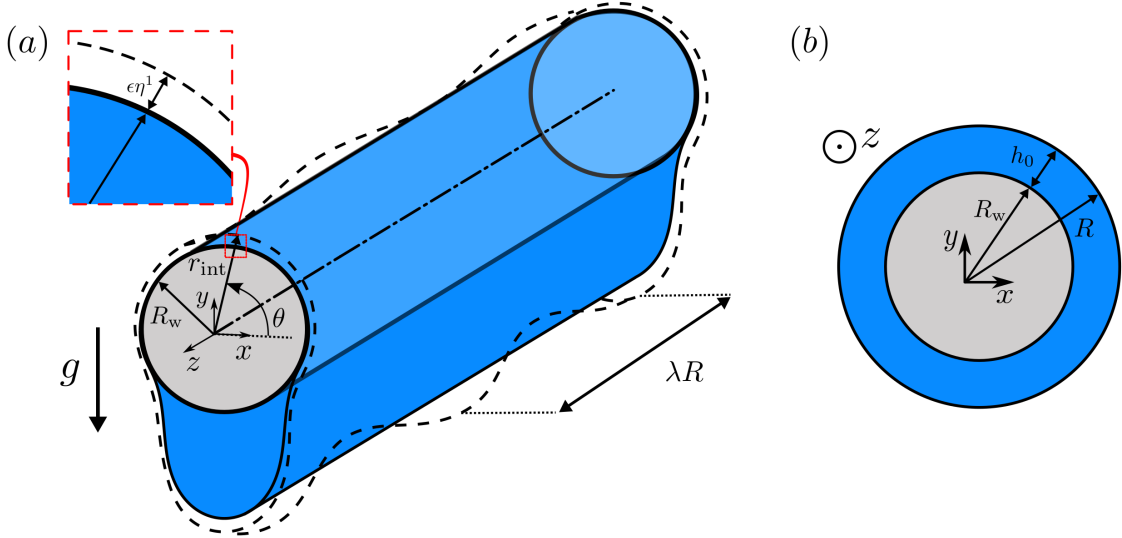}}% Images in 100% size
	\caption{(a) Schematic of the liquid film coating the outside of a horizontal cylinder and the geometrical parameters. The thick solid black line shows the cylinder wall of the radius $R_\text{w}$, centred at $(0,0,0)$. The liquid-gas interface is shown in the thin solid black line. Grey colour marks the interior cross-section of the solid cylinder. The dashed black line represents the perturbed liquid-gas interface of the local radius $r_\text{int}$ and axial wavelength $\lambda R$. The inset shows the zoomed cross-section of the perturbed interface. Gravity acts vertically, perpendicular to the tube axis; (b) The initial $x-y$ cross-section of the liquid column.} 
	\label{fig:horizontalfiber_schematic}
\end{figure}

The outer wall of a rigid and immobile solid circular cylinder of radius $R_\text{w}$ is coated with a viscous liquid film. The schematic of the flow is presented in figure~\ref{fig:horizontalfiber_schematic}(a). The standard Cartesian coordinates $(x,y,z)$ are considered with the origin placed at the center of the solid cylinder. In-plane coordinates are $(x,y)$, and the gravity acceleration ${\bf g}$, points in the $-y$ direction.
We consider a Newtonian liquid of constant dynamic viscosity $\mu$, surface tension $\gamma$, and density $\rho$, surrounded by an inviscid immobile gas.
The interface radius $r_\text{int}$ is parametrised in cylindrical coordinates $(r,\theta,z)$ as $\mathcal{F} = r-r_\text{int}(t,\theta,z)=0$, using the same origin as the Cartesian one.
The liquid-gas interface is initially concentric with the cylinder and the liquid film thickness is constant around the periphery of the wall, $h_0=R-R_\text{w}$, where $R$ denotes the initial interface radius (figure~\ref{fig:horizontalfiber_schematic}(b)).
At dimensionless time $t=0$, the initial condition writes $r_\text{int}(0,\theta,z)=R$.
The dimensionless state vector ${\bf q}= ({\bf u},p,\Rint)^T$ describes the liquid motion at any instance $t$, where~${\bf u}(t,x,y,z)= (u_x,u_y,u_z)^T$ denotes the three-dimensional velocity field, $p(t,x,y,z)$ denotes the pressure, and $\Rint=r_\text{int} / R$ denotes the dimensionless interface radius.
We choose the initial interface radius $R$ as the length scale and its static pressure jump as the pressure scale.
The intrinsic velocity scale associated with a viscous liquid film of thickness $h_0$ falling under its weight, presented by~\citet{Duclaux2006}, is chosen to make the state vector and the governing equations dimensionless.
Additionally, the advection time scale is constructed based on the aforementioned velocity and length scales as:

\begin{equation} \label{eq:horizontalfiber-Gauges}
	\begin{aligned}
		\mathcal{L}&=R , & \mathcal{U}&=\frac{\rho g h_0^2}{\mu}=\frac{\rho g R^2}{\mu} (1-\beta)^2,\\
		\mathcal{P}& =\frac{\gamma}{R}, &\mathcal{T}&=\frac{\mathcal{L}}{\mathcal{U}}= \frac{\mu}{\rho g R} (1-\beta)^{-2},
	\end{aligned}
\end{equation}
where $\beta= {R_\text{w}}/{R}$ denotes the dimensionless wall radius. Consequently, the dimensionless value of the initial film thickness is expressed as $\delta = h_0 / R = 1-\beta$. The flow is governed by the conservation of mass and momentum equations which in dimensionless form read

\begin{equation} \label{eq:horizontalfiber-Incompressibility}
	\nabla \cdot \,{\bf u}=0,
\end{equation}
\begin{equation} \label{eq:horizontalfiber-Navier-Stokes-dimless}
	\left(\frac{Bo}{Oh}\right)^2 \delta^4 \left( \p_t + {{\bf u}}\cdot \nabla \right) {\bf u} =\nabla \cdot \, \stau - Bo \text{ \bf e}_y,
\end{equation}
respectively, where $\p_j$ denotes the partial derivative with respect to quantity $j$, and the stress tensor \(\underline{\underline{\tau}}\) is expressed as
\begin{equation} \label{eq:horizontalfiber-Stresstensor-dimless}
	\stau=-\, {p} \tens{I} + Bo \ \delta^2 \left( \nabla {{\bf u}} + \nabla {{\bf u}}^T \right). 
\end{equation}
%%%%%
Two other dimensionless numbers appear in the governing equations: the \textit{Ohnesorge} number, $Oh={\mu}/{\sqrt{\rho \gamma R}}$, compares the viscous and inertial forces. The \textit{Bond} number, $Bo={\rho g R^2}/{\gamma}$, compares the gravitational and surface tension forces. Our study addresses the limit of inertialess flow where $\left( {Bo}/{Oh} \right)^2 \delta^4 \ll 1$. One can re-express $\left( {Bo}/{Oh} \right)^2 \delta^4$ as $ Re \, \delta^2$ where the \textit{Reynolds} number, $Re = {\rho \mathcal{U} \mathcal{L}}/{\mu}$, is constructed upon the same scales presented in~(\ref{eq:horizontalfiber-Gauges}).

The no-slip boundary condition, $\bf u = 0$, is applied on the solid wall, at $r=\beta$. On the shear-free fluid-gas interface, the kinematic and dynamic boundary conditions write

\begin{equation} \label{eq:horizontalfiber-kinematic-BC}
	\p_t \Rint + {\bf u} \cdot \nabla \Rint= {\bf u \cdot e}_r \quad \text{at } \ r=\Rint,
\end{equation}

\begin{equation} \label{eq:horizontalfiber-dynamic-BC}
	\underline{\underline{{\tau}}} \cdot  {{\bf n}} = - \kappa {{\bf n}}  \quad \text{at } \ r=\Rint,
\end{equation}
respectively, where ${\bf e}_r$ denotes the unit radial vector, ${\bf n} = \nabla \left( r-\Rint\right) / \left\Vert \nabla \left( r-\Rint\right) \right\Vert$, denotes the interface unit normal vector pointing outward from the origin of the coordinate system, $\left\Vert \cdot \right\Vert$ denotes the Euclidean norm, and $\kappa=\nabla \cdot {{\bf n}}$ denotes the mean curvature of the interface.
%%%%%%%%%
%
To build intuition about the flow characteristics and before describing the stability analysis and numerical method, we illustrate the reference flow in \S\ref{sec:horizontalfiber-baseflow}.

\subsection{Base flow} \label{sec:horizontalfiber-baseflow}

\begin{figure}
	\centerline{\includegraphics[width=1\textwidth]{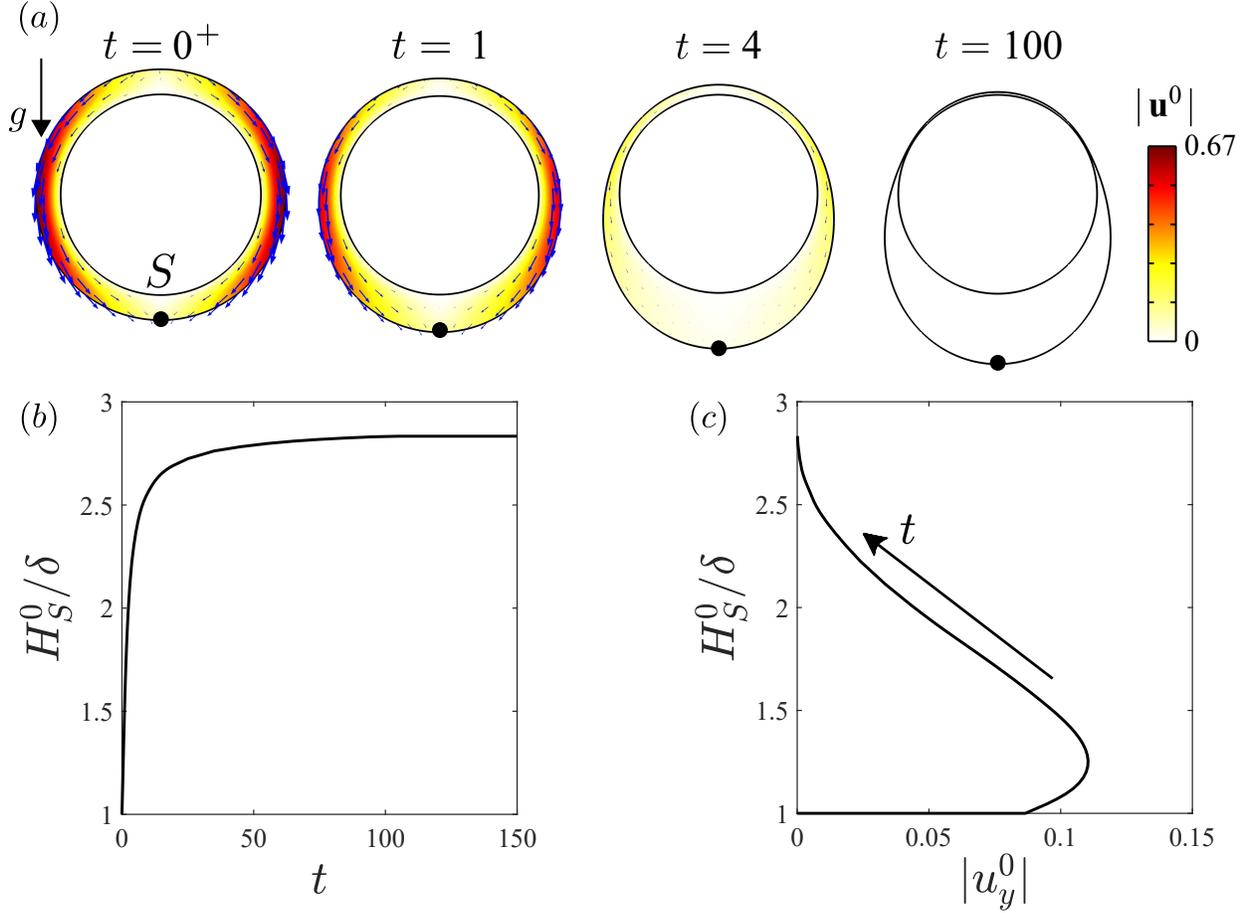}} 
	\caption{Base flow evolution for $Oh \rightarrow \infty, Bo=0.4, \delta =0.2$. (a) Snapshots of the flow field: colour map shows the in velocity magnitude, arrows show the liquid velocity field, and the point $S$ marks the south pole of the bubble, $\theta_S=3\pi/2$.
		(b) Temporal variation of the relative liquid film thickness at the pole $H_S^0/\delta = (\Rb(\theta_S)-\beta)/\delta$.
		(c) Film thickness variation as a function of the vertical velocity at the pole. }	\label{fig:horizontalfiber-base-flow}
\end{figure}

\begin{figure}
	\centerline{\includegraphics[width=1\textwidth]{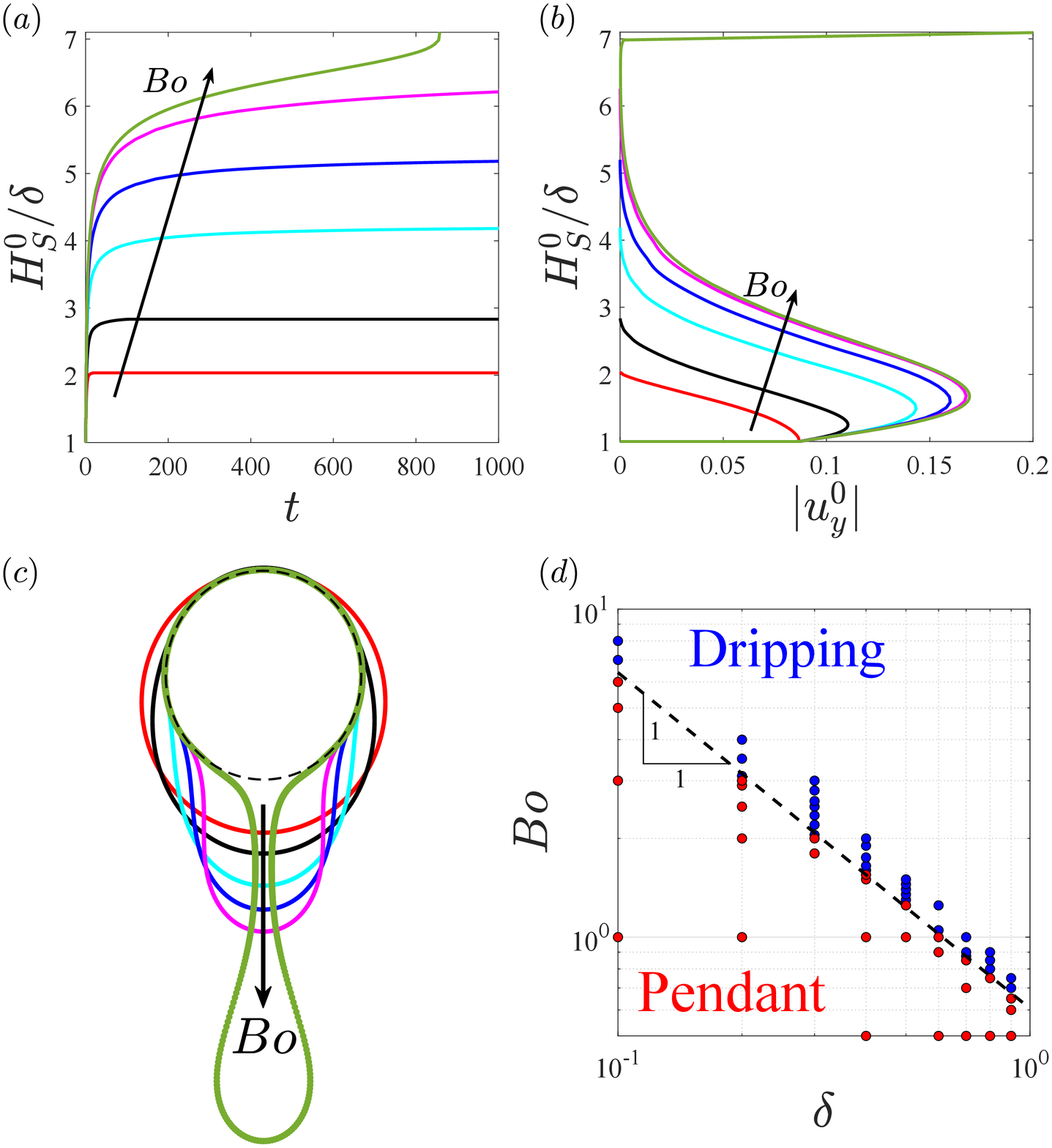}} 
	\caption{Influence of the surface tension. The base flow dynamics for $Oh \rightarrow \infty, \delta =0.2, Bo=\{ 0.01, 0.4,1.6,2.5,3,3.1\}$: (a) Temporal variation of the relative liquid film thickness at the pole. (b) Film thickness variation as a function of the vertical velocity at the pole. (c) The large-time interfaces of the pendant curtains at $Bo=\{ 0.01, 0.4,1.6,2.5,3\}$, and the dripping curtain at $Bo=3.1$ (green) for the same flows presented in panels (a-b). The black dashed circle shows the cylinder. (d) Two-dimensional pendant vs dripping diagram in the $\delta-Bo$ plane: the black dashed line corresponds to the best fit to the two-dimensional pendant to dripping transition: $Bo=0.61 \delta^{-1}$.}	\label{fig:horizontalfiber-base-flow-Bo}
\end{figure}

The {\it base flow}, denoted by ${\bf q}^0$, is the two-dimensional, in-plane, transient solution of the non-linear conservation equations~(\ref{eq:horizontalfiber-Incompressibility})-(\ref{eq:horizontalfiber-dynamic-BC}) where the fluid is initially assumed at rest with constant initial pressure $p^0(t=0)=1 $ and $\Rb=1$.
Due to the non-linear nature of the interface conditions, finding an analytical solution for such a flow is challenging. Hence, the temporal evolution of the flow is computed numerically (see \S\ref{sec:horizontalfiber_Numerical-method} for details).
%%%
Some snapshots from the base flow evolution are shown in figure~\ref{fig:horizontalfiber-base-flow} for an exemplary case of $\delta=0.2$, and an intermediate Bond number $Bo=0.4$.
The dynamics of the base flow, presented in figure~\ref{fig:horizontalfiber-base-flow}(b-c), can be characterised by quantifying the relative liquid film thickness $H^0/\delta = (\Rb-\beta)/\delta$ at the south pole cylinder, $\theta=3\pi /2$, and its vertical velocity.
This point is hereafter referred to as the pole, where the strongest gravitational effects are expected.
Drainage begins with an immediate liquid pull-off around the solid cylinder that decays with time. The liquid body forms a two-dimensional quasi-static pendant curtain as $t\rightarrow \infty$. \\
Drainage dynamics are significantly influenced by surface tension and gravity effects, i.e. by $Bo$.
Figure~\ref{fig:horizontalfiber-base-flow-Bo} presents the influence of the Bond number.
By increasing $Bo$, i.e. weakening the effect of surface tension against gravity, flow initially accelerates before decaying and the formation of the quasi-static curtain gets delayed (figure~\ref{fig:horizontalfiber-base-flow-Bo}(a)).
Exceeding a critical $Bo$ value, the interface cannot sustain the liquid weight anymore and the pole accelerates downward sharply (green line in figure~\ref{fig:horizontalfiber-base-flow-Bo}(a-b)), thus causing a two-dimensional rupture.
Figure~\ref{fig:horizontalfiber-base-flow-Bo}(c) shows the interface of the quasi-static pendant curtain for several $Bo$ values as well as one example of dripping bulk after re-accelerating.
While small $Bo$ numbers sustain the quasi-circular cross-section of the curtain, increasing $Bo$ results in a stretched interface in the direction of gravity.
The green line in figure~\ref{fig:horizontalfiber-base-flow-Bo}(c) depicts the falling liquid interface right after the pole acceleration (at $t=855.8$)  for $Bo=3.1$.
Figure~\ref{fig:horizontalfiber-base-flow-Bo}(d) presents the region of saturation towards a static pendant curtain in the $\delta-Bo$ plane.
This diagram is obtained by following the non-linear drainage simulations until the maximal flow velocity magnitude drops below $10^{-5}$ (pendant) or the occurrence of the large-time pole acceleration (dripping). It was verified for some marginal sub-critical pairs of $\{Bo,\delta\}$ that lowering the velocity threshold down to $10^{-7}$ does not affect the critical parameters at the transition between these two behaviours.
The dripping occurs if $Bo > 0.61 \delta^{-1}$, suggested by the best fit to the data from our simulations.

We now follow a scaling argument to rationalise the obtained threshold. In the case of a pendant curtain, the capillary force per unit axial length $\propto 2 \gamma$ overcomes the bulk weight, $\pi \rho g R^2 \delta (1+\beta)$. This comparison implies the dripping threshold as $Bo > (2 / \pi) \delta^{-1} (1+\beta)^{-1}$ where $2 / \pi \approx 0.64$. 
In the thick-film limit, $\beta \ll 1 $ (remember that $\delta=1-\beta$), the dripping threshold approaches $0.64 \delta^{-1}$ which is in good agreement with our numerics.
Following this scaling argument, by approaching the thin-film limit, $\beta \rightarrow 1$, the dripping threshold should approach $Bo > 0.32 \delta^{-1}$. However, this scaling contradicts the numerical observation.
One might propose a correction to this scaling argument by taking into account the tangential direction at which the minimum capillary force is applied.
However, such a correction was found unsuccessful, too.
Note that rupture occurs through the pole acceleration, and necking takes place at some distance below the cylinder, where surface tension fails to withstand merely the portion of the liquid weight that accelerates underneath the neck (see the green interface in figure~\ref{fig:horizontalfiber-base-flow-Bo}(c)).

\subsection{Linear stability analysis of the pendant curtain} \label{sec:horizontalfiber-formulation-stability-analysis}

To conduct the {\it linear stability analysis} of the quasi-static pendant curtain, presented in \S\ref{sec:horizontalfiber-baseflow}, the state vector ${\bf q}= ({\bf u},p,\Rint)^T$ is decomposed into the sum of the steady saturated base flow solution ${\bf q}_\infty^0$ (subscript $_\infty$ denotes large-time evaluation), and the infinitesimal time-dependent {\it perturbation} ${\bf q}^1= \left( {\bf u}^1,p^1,\eta^1 \right)^T$, i.e.
%%%%%%%%%%
% 
\begin{equation} \label{eq:horizontalfiber-Solution-decomposition}
	{\bf q}={\bf q}_\infty^0 + \epsilon {\bf q}^1 + \mathcal{O}(\epsilon^2), \quad \epsilon \ll 1,
\end{equation}
where the amplitude $\epsilon$ is small.
The normal mode of the perturbation ${\bf q^1}$ with the longitudinal wavenumber $k$ (associated with the wavelength $\lambda={2 \pi}/{k}$) reads
\begin{equation} \label{eq:horizontalfiber_eigenmode_ansatz}
	{\bf q}^1=\tilde{{\bf q}}(x,y) \ \mathrm{exp} \left[ \sigma t+\mathrm{i}kz \right] + \mathrm{c.c.},
\end{equation} 
where c.c. denotes the complex conjugate.
All other functions in terms of the state vector can be decomposed in a similar fashion.
Namely, $ \stau = \stau^0+ \epsilon \stau^1 $, $ {\bf n}=  {\bf n}^0+ \epsilon {\bf n}^1 $ and $ \kappa=  \kappa^0+ \epsilon \kappa^1 $. (For further details about the formulation of ${\bf n}^1$ and $\kappa^1$, see
Appendix C in \citet{Eghbali2022in}.)
In the asymptotic limit of large times, a normal eigenmode perturbation with complex eigenvalue $\sigma=\sigma_r + \mathrm{i} \sigma_i$ is {\it unstable} if $\sigma_r > 0$.
An unstable eigenmode grows exponentially in time with the growth rate $\sigma_r$. (Unless otherwise noted, the indices $r$ and $i$ denote the real and imaginary parts of a complex number, respectively.)
After casting the perturbed state of~(\ref{eq:horizontalfiber-Solution-decomposition}) into the governing equations~(\ref{eq:horizontalfiber-Incompressibility})-(\ref{eq:horizontalfiber-Navier-Stokes-dimless}), with the static pendant curtain base state ${\bf q}_\infty^{0}=\left( {\bf u}^0,p^0,\Rb \right)^T$, and neglecting the higher orders in $\epsilon$, the linearised equations can be expressed as
\begin{equation} \label{eq:horizontalfiber-linearised-Incompressibility}
	\nabla\cdot\,{\bf u}^1=0,
\end{equation}
\begin{equation} \label{eq:horizontalfiber-linearised-momentum}
	\left(\frac{Bo}{Oh}\right)^2 \delta^4 \left( \p_t {\bf u}^1 + \left({\bf u}^0 \cdot \nabla \right) {\bf u}^1 + \left( {\bf u}^1 \cdot \nabla \right) {\bf u}^0  \right) = \nabla \cdot\,\underline{\underline{\tau}}^1.
\end{equation}
The corresponding boundary conditions are as follows. On the solid cylinder boundary, $r=\beta$, the no-slip condition implies ${\bf u}^1={\bf 0}$ (yielding ${\bf \tilde{u}=0}$).
As the geometry of the perturbed flow is unknown, the interface conditions~(\ref{eq:horizontalfiber-kinematic-BC})-(\ref{eq:horizontalfiber-dynamic-BC}), applied on the perturbed liquid interface $r=\Rb + \epsilon \eta^1$, should be projected radially onto the base interface, $r=\Rb$, and ultimately linearised; a process called {\it flattening} (see~(\ref{eq:horizontalfiber-flattening}) in appendix~\ref{app:horizontalfiber_interfacebc}).
The kinematic condition once linearised implies

\begin{equation} \label{eq:horizontalfiber_Interface_BC_linearised_kin}
	\p_t {\eta}^1 + \underbrace{  \left(  -\p_r u^0_r + \frac{\p_r u^0_\theta \ \p_\theta \Rb}{\Rb} - \frac{u^0_\theta \ \p_\theta \Rb }{\left(\Rb\right)^2} \right) {\eta}^1 + \frac{u^0_\theta  }{\Rb} \p_\theta {\eta}^1 }_{-\tens{G}^0 {\eta}^1} + \ \frac{\p_\theta \Rb}{\Rb}  {u}_{\theta}^1 = {{u}_r^1 }, \quad \text{at} \ r=\Rb,
\end{equation}

where $\left( u^0_r, u^0_\theta,0 \right)^T $ and $\left( {u}_r^1, {u}_\theta^1, {u}_z^1 \right)^T $ denote the velocity vectors of the base state and perturbations, respectively, represented in the cylindrical coordinates.
Even though in the case of a quasi-static pendant curtain, ${\bf u}^0 \approx {\bf 0}$, we keep the corresponding terms in the linearised equations (\ref{eq:horizontalfiber_Interface_BC_linearised_kin}) and the following equations.
Introducing a normal eigenmode of the form~(\ref{eq:horizontalfiber_eigenmode_ansatz}) into~(\ref{eq:horizontalfiber-linearised-Incompressibility})-(\ref{eq:horizontalfiber-linearised-momentum}), combined with~(\ref{eq:horizontalfiber_Interface_BC_linearised_kin}), leads to a generalised eigenvalue problem for $\sigma$ and $\tilde{{\bf q}}$ as

\begin{equation} \label{eq:horizontalfiber-eigenvalue-problem}
	\tens{L} \tilde{{\bf q}} + \text{c.c.} =\sigma \tens{B} \tilde{{\bf q}} + \text{c.c.},
\end{equation}
where the linear operators $\tens{L}$ and $\tens{B}$ can be expressed as
% \begin{align} \label{eq:horizontalfiber_eigenvalue_operators_global}
% 	\nonumber \tens{L} & = \begin{bmatrix}
% 		\left(\frac{Bo}{Oh}\right)^2 \delta^4 \tens{F}^0 + Bo \delta^2 \left( \tilde{\nabla}\cdot(\tilde{\nabla}+\tilde{\nabla}^T) \right)       & -\tilde{\nabla} & {\bf 0} \\
% 		\tilde{\nabla}\cdot       & 0 & 0 \\
% 		\left( {\bf e}_r - \frac{\p_\theta \Rb}{\Rb} {\bf e}_\theta \right) \cdot       & 0 & \tens{G}^0
% 	\end{bmatrix}, \\  
% 	\tens{B}& = \begin{bmatrix}
% 		\left(	\frac{Bo}{Oh} \right)^2 \delta^4 \tens{I}   & {\bf 0} & {\bf 0}  \\
% 		{\bf 0} & 0 & 0 \\
% 		{\bf 0} & 0 & 1 \,  
% 	\end{bmatrix},
% \end{align}
\begin{align} \label{eq:horizontalfiber_eigenvalue_operators_global}
	\nonumber \tens{L} & = \begin{bmatrix}
		\left(\frac{Bo}{Oh}\right)^2 \delta^4 \tens{F}^0 + Bo \delta^2 \tens{J}& -\tilde{\nabla} & {\bf 0} \\
		\tilde{\nabla}\cdot       & 0 & 0 \\
		\left( {\bf e}_r - \frac{\p_\theta \Rb}{\Rb} {\bf e}_\theta \right) \cdot       & 0 & \tens{G}^0
	\end{bmatrix}, \\  
	\tens{B}& = \begin{bmatrix}
		\left(	\frac{Bo}{Oh} \right)^2 \delta^4 \tens{I}   & {\bf 0} & {\bf 0}  \\
		{\bf 0} & 0 & 0 \\
		{\bf 0} & 0 & 1 \,  
	\end{bmatrix},
\end{align}
where $\tens{F}^0 \tilde{\bf u} = -\bigl( \left({\bf u}^0 \cdot \tilde{\nabla} \right) \tilde{\bf u} + \left( \tilde{\bf u} \cdot \nabla \right) {\bf u}^0 \bigr) $ and $\tens{J} \tilde{\bf u} = \tilde{\nabla} \cdot (\tilde{\nabla} \tilde{\bf u} + (\tilde{\nabla} \tilde{\bf u})^T)$.
Here, $({ \bf e}_r,{ \bf e}_{\theta},{ \bf e}_z)$ denote the unit direction vectors in the cylindrical coordinates $(r,\theta,z)$ used for parameterising the interface, and the gradient operators and the velocity gradient tensors in the Cartesian coordinates read

\begin{align} \label{eq:horizontalfiber-gradient-operator-global}
	\nonumber	{\nabla} & = (\p_x, \p_y,\p_z)^T, &{\nabla} {\bf u}^0 &= \begin{bmatrix}
		\p_x {u}^0_x & \p_y {u}^0_x & 0  \\
		\p_x {u}^0_{y} & \p_y {u}^0_{y} & 0  \\
		0 & 0 & 0\, 
	\end{bmatrix}, \\
	\tilde{\nabla} & = (\p_x, \p_y,\mathrm{i}k)^T,
	&\tilde{\nabla} \tilde{{\bf u}} &= \begin{bmatrix}
		\p_x \tilde{u}_x & \p_y \tilde{u}_x & \mathrm{i}k \tilde{u}_x  \\
		\p_x \tilde{u}_{y} & \p_y \tilde{u}_{y} & \mathrm{i}k \tilde{u}_{y}  \\
		\p_x \tilde{u}_z & \p_y \tilde{u}_z & \mathrm{i}k \tilde{u}_z  \, 
	\end{bmatrix}.
\end{align}
%%%%%%%%%%%%
The interface dynamic condition~(\ref{eq:horizontalfiber-dynamic-BC}), once linearised and considering the normal mode \eqref{eq:horizontalfiber_eigenmode_ansatz}, reads
%%%%%%%%%%%%
\begin{equation} \label{eq:horizontalfiber_Interface_BC_linearised_dyn}
	\underline{\underline{{\tau}}}^0 \cdot \tilde{\bf n} + \tilde{\eta} \, \p_r \underline{\underline{{\tau}}}^0 \cdot {\bf{n}}^0 + \underline{\underline{{\tilde{\tau}}}} \cdot {\textbf{n}}^0  =  -\left( \kappa^0 \tilde{\bf n} + \tilde{\kappa} {\textbf{n}}^0 \right),  \quad \text{at} \ r=\Rb.
\end{equation}
This condition is imposed while solving the system \eqref{eq:horizontalfiber-eigenvalue-problem} numerically. (For further details on the derivation of the interface conditions and their numerical implementation, see appendices~\ref{app:horizontalfiber_interfacebc} and~\ref{app:horizontalfiber_implementation_LSA}, respectively.)

\subsection{Numerical method} \label{sec:horizontalfiber_Numerical-method}
\begin{figure}
	\centerline{\includegraphics[width=0.3\textwidth]{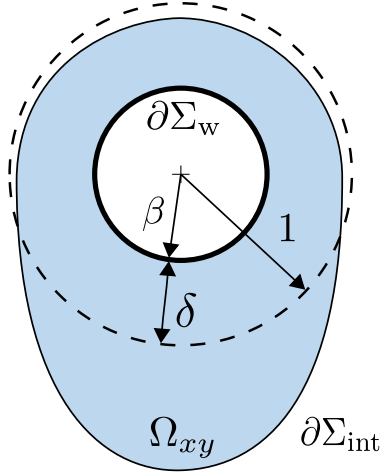}	}% Images in 100% size
	\caption{\label{fig:horizontalfiber_numerical_domain_drainage_cartesian} The numerical domain for computing the base flow
 %, transient growth study,
    and linear stability analysis. Here, $\Omega_{xy}$ denotes the liquid bulk. The boundaries of the numerical domain are denoted by $\p \Omega_{xy} = \p \Sigma_{\mathrm{w}} \cup \p \Sigma_{\mathrm{int}}$, where $\p \Sigma_{\mathrm{w}}$ represents the exterior wall of the cylinder with the radius of $\beta$, and $\p \Sigma_{\mathrm{int}}$ represents the gas-liquid interface. The cross-section of the interface is initially a circle of unit radius, concentric with the cylinder (sketched in black dashed line).}
\end{figure}
The base flow
%, transient growth, 
and linear stability analyses are carried out numerically by the finite element software \comsol. A triangular moving mesh is generated on the two-dimensional domain shown in figure~\ref{fig:horizontalfiber_numerical_domain_drainage_cartesian}. The grid size is controlled by the vertex densities on the boundaries $\p \Sigma_{\mathrm{w}}$ and $\p  \Sigma_{\mathrm{int}}$. The variational formulation of the base flow equations~(\ref{eq:horizontalfiber-Incompressibility})-(\ref{eq:horizontalfiber-dynamic-BC}), linear stability equations (\ref{eq:horizontalfiber-eigenvalue-problem}), and linearised Navier-Stokes equations (\ref{eq:horizontalfiber-tg-ode}) are discretised spatially using quadratic (P2) Lagrange elements for the geometrical shape function, ${\bf u}^0$, $\tilde{{\bf u}}$, ${\bf \bar{u} }$, $\tilde{\eta}$ and $\bn$, and linear (P1) Lagrange elements for $p^0$, $\tilde{p}$, and $\bar{p}$. This discretisation results in approximately $N_{dof}=$ 400'000 degrees of freedom for the base flow calculations as well as the linear stability analyses.

First, the base flow is computed using the laminar two-phase flow module incorporated with the moving mesh module. The numerical time step is determined by the Backward differentiation formula with maximum differentiation order of 2. The solver is initialised by the Backward Euler consistent initialisation with an initial step fraction of $10^{-9}$. At each time step, Newton's method is used to solve the non-linear equations, where the relative tolerance for the iterative solver convergence is set to $10^{-6}$.
In the built-in module of \comsol, the kinematic condition (\ref{eq:horizontalfiber-kinematic-BC}) is replaced by its equivalent form and readily implemented enforcing ${\bf u} \cdot {\bf n} =	{\bf u}_{mesh} \cdot {\bf n}$, at $\p \Sigma_\text{int}$, where ${\bf u}_{mesh}$ denotes the moving mesh velocity.
Following the computed base flow, the first solution after the maximal velocity falls below $10^{-5}$ is considered as the static pendant curtain. 
Then the linear stability analysis is conducted by solving the generalised eigenvalue problem~(\ref{eq:horizontalfiber-eigenvalue-problem}) for the static pendant curtain using the shift-invert Arnoldi method.
(For more details on the development of the variational formulation, implementation of the linearised Navier-Stokes and linear stability eigenvalue problem, and their corresponding boundary conditions see appendix~\ref{app:horizontalfiber_implementation_LSA}.) 
The computation time for obtaining the base flow for a given set of parameters followed by the stability analysis for $\sim$20 values of $k$ is of the order of an hour on a single Intel core at 3.6 GHz. 
Both the base flow and stability analysis models are validated with the existing solutions in the literature. (For more details about the series of validation tests see appendix~\ref{app:horizontalfiber_validation}.)

\section{Results} \label{sec:horizontalfiber_results}
The results of the linear stability analysis
% and transient growth analysis 
are presented hereafter. To begin with, an overview of the stability of the pendant curtain and the influence of $Bo$ are presented in \S\ref{sec:horizontalfiber_results_LSA_overview}.
Then both the base flow and the perturbations feeding on the pendant curtain are further discussed from an energy viewpoint in \S\ref{sec:horizontalfiber_res_energy}, enabling us to investigate the contributions of different physical mechanisms in the formation of the pendant curtain and its instability and to complement the stability diagram by identifying regions where either the Rayleigh-Plateau or the Rayleigh-Taylor instabilities dominate.
In \S\ref{sec:horizontalfiber_results_3dpinch}, the most asymptotically amplified linear modes and the consequent pattern formation are discussed, followed by sketching a provisional regime diagram of the pendant curtain predicted by the linear analysis.
\subsection{Linear stability of the pendant curtain} \label{sec:horizontalfiber_results_LSA_overview}
%%

% \subsubsection{Effect of surface tension ($Bo$)} \label{sec:horizontalfiber_results_param}
\begin{figure}
	\centerline{\includegraphics[width=1\textwidth]{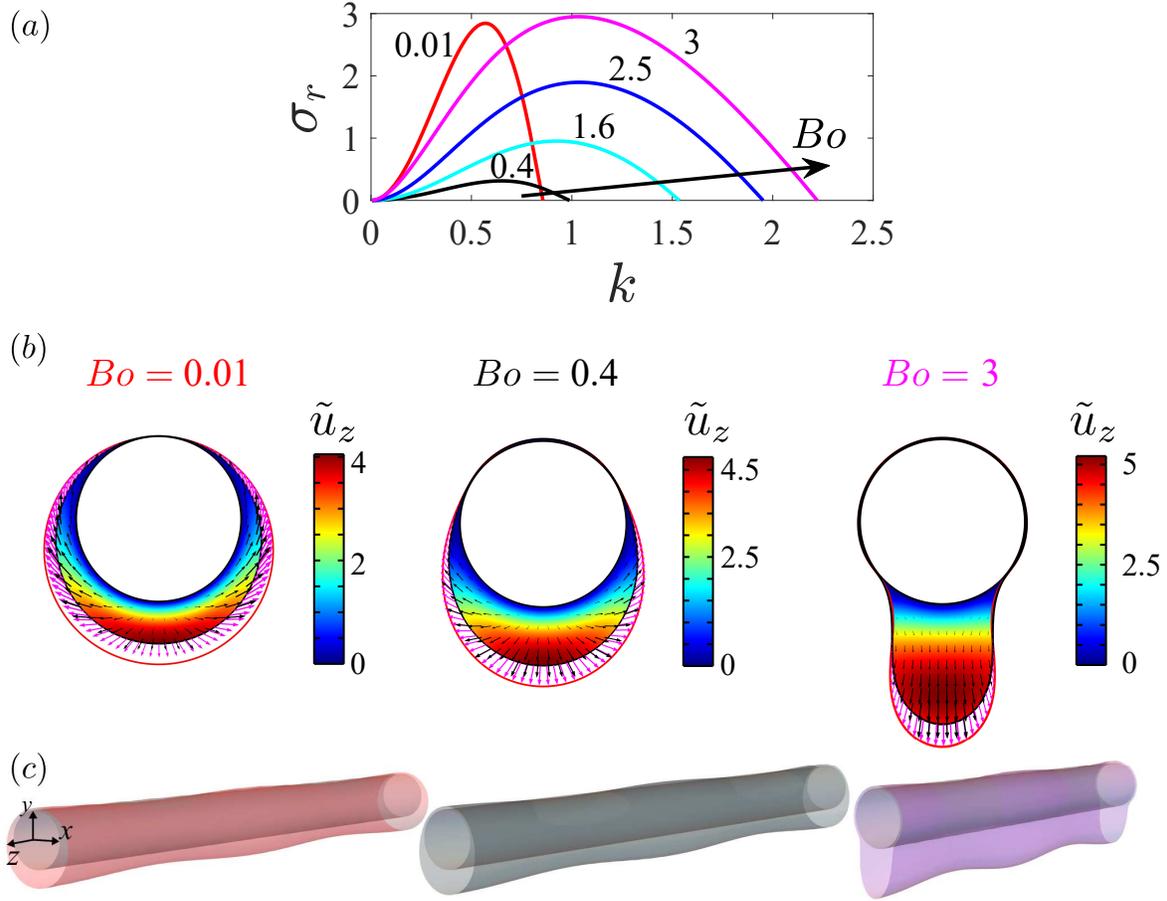}} 
	\caption{The influence of $Bo$ on the linear stability of the pendant curtains whose base flows are presented in figure~\ref{fig:horizontalfiber-base-flow-Bo}:
		(a) Dispersion curve of the unstable eigenmode;
		(b) Eigenvelocity fields and eigeninterfaces of the same unstable modes at $k_{max}$, corresponding to $\smax$; the color map presents the axial eigenvelocity, black arrows show the in-plane eigenvelocity field; magenta arrows show the eigenvelocity at the base interface, and the red line renders the base interface perturbed by an arbitrary amplitude;
		(c) The three-dimensional render of the asymptotically most amplified perturbed interfaces, shown in panel (b);
		$Oh \rightarrow \infty, \ \delta =0.2$.}
	\label{fig:horizontalfiber_LSA_overview}
\end{figure}
%%%%%%
In this section, we present the linear stability characteristics associated with the quasi-static pendant curtain coating the outside of a horizontal cylinder.
We follow the same exemplary cases of $\delta=0.2$ the base flows of which are presented in \S\ref{sec:horizontalfiber-baseflow}.
We found that each set of parameters exhibits only a single unstable mode whose {\it dispersion curve}, representing the growth rate $\sigma_r(k)$, is presented in figure~\ref{fig:horizontalfiber_LSA_overview}(a).
This mode is unstable within a range of wavenumbers $0 \le k \le k_c$, where $k_c$ denotes the cut-off wavenumber, and exhibits a peak in its growth rate, $\smax$, at an associated maximal mavenumber $\kmax$.
Starting from large surface tension compared to gravity, $Bo=0.01$, increasing $Bo$ results first in a decrease in $\smax$ (from $Bo=0.01$ to $Bo=0.4$) followed by a rebound for large $Bo$ (for $Bo>0.4$).
The cut-off wavenumber $k_c$ increases monotonously, whereas $\kmax$ increases up to a saturated value $k \approx 1.05$.

The perturbation eigenvelocity fields at $\kmax$ and three-dimensional rendering of the perturbed interface with an arbitrary amplitude are presented for $Bo=\{ 0.01,0.4,3\}$ in figure~\ref{fig:horizontalfiber_LSA_overview}(b-c), respectively.
Each eigenstate $\tilde{q}$ is normalised with its RMS value, and its phase is set such that the axial velocity of the pole, $\tilde{u}_z(\theta=-\pi/2)$, becomes real-valued.
The unstable mode features left/right symmetry, strong interface modulation at the bottom of the pendant curtain, and an immobile interface at the top of the cylinder, $\theta=\pi/2$.
While intermediate and strong surface tension to gravity ratios, $Bo=\{0.01,0.4\}$, evidence the flow reversal towards the top side of the cylinder, when gravity dominates at $Bo=3$, interface perturbations take place only at the bottom of the pendant curtain, promoting vertical fingers underneath the cylinder. 
Similar patterns were reported through the non-linear simulations of~\citet{Weidner1997} in the thin film limit with large surface tension.
For one particular case of liquid roll-up, they observed that the non-linear evolution of the perturbations results in lower surface energy at the cost of increasing the potential energy of the liquid.
On the contrary, for one case of gravity dominance, the non-linear perturbation evolution was reported in favour of the potential energy reduction despite increasing the surface energy.
These arguments suggest further study of the base flow and linear perturbations from an energy perspective to quantify the effect of different physical mechanisms at play in the flow.
%%%%%

\subsection{Energy analysis} \label{sec:horizontalfiber_res_energy}
In this section, we study the flow from an energy point of view in order to clarify the interactions between capillary, viscous, and gravitational effects, and to quantify their respective contributions to the formation and linear instability of a pendant curtain. Formerly,~\citet{Hooper1983,Boomkamp1996,Kataoka1997,Li2011} employed this method to evaluate and compare the role of different physical mechanisms on the temporal instability of various interfacial flows.
%%%%%%
Hereafter, the area increment in the bulk cross-section is denoted by $ \mathrm{d} A_{\Omega_{xy}}$. On the boundary $j$, the increment of the surface area is denoted by $ \mathrm{d} A_{\Sigma_j}$, and the increment of the arc length is denoted by $ \mathrm{d} s$.

Here, we study energy conservation in the flow, on different scales, from the base flow to the perturbations. We focus on the base flow presented in \S\ref{sec:horizontalfiber-baseflow} and the unstable modes presented in \S\ref{sec:horizontalfiber_results_LSA_overview}. More precisely, the energy analysis sheds light on the balance of the energy rate, hereafter referred to as the {\it energy equation}, which for the inertialess gravity-driven flow down a horizontal cylinder can be expressed as
\begin{equation} 
	\underbrace{ \iiint_{\Omega_{xy}} Bo \ \delta^2 \ tr\left(  \left( \nabla {\bf u} + (\nabla {\bf u})^T \right) \cdot \nabla {\bf u } \right) \mathrm{d} V}_{\text{DIS}} + \underbrace{ \iint_{\p \Sigma_{\mathrm{int}}} -\left( \underline{\underline{{\tau}}} \cdot  {\bf n}^0 \right) \cdot {\bf u} \ \mathrm{d} A_{\Sigma_\text{int}}}_{\text{BND} } + \underbrace{ \iiint_{\Omega_{xy}}  - Bo \ u_y \ \mathrm{d} V}_{\text{POT}} = 0,
	\label{eq:horizontalfiber_energy_balance_dimless_simp}
\end{equation}
%%%%
where the bulk integral is defined on the volume increment $\mathrm{d} V=\dAb  \mathrm{d} z$, the surface integral is defined on the columnar surface with the cross-section $\p \Sigma_{\mathrm{int}}$ and axis in $z$ direction (see figure~\ref{fig:horizontalfiber_numerical_domain_drainage_cartesian}), DIS denotes the rate of viscous dissipation in the bulk fluid, BND denotes the rate of interfacial work conducted by the fluid, and POT denotes the rate of change of gravitational potential energy. (For more details about the derivation of the energy equation, see appendix~\ref{app:horizontalfiber_energyderiv}.)
The energy equation implies that the net rate of energy exchange in the flow is zero, where multiple physical mechanisms may contribute to energy release and consumption. The sign of each term in~(\ref{eq:horizontalfiber_energy_balance_dimless_simp}) indicates whether the energy is removed from ($+$) or released into ($-$) the flow by the respective mechanism.
%%%%%%%
%%%%%%%
\subsubsection{Energy analysis of the base flow} \label{sec:horizontalfiber_res_energy_Baseflow}
%%%%%%%
\begin{figure}
	\centerline{\includegraphics[width=1\textwidth]{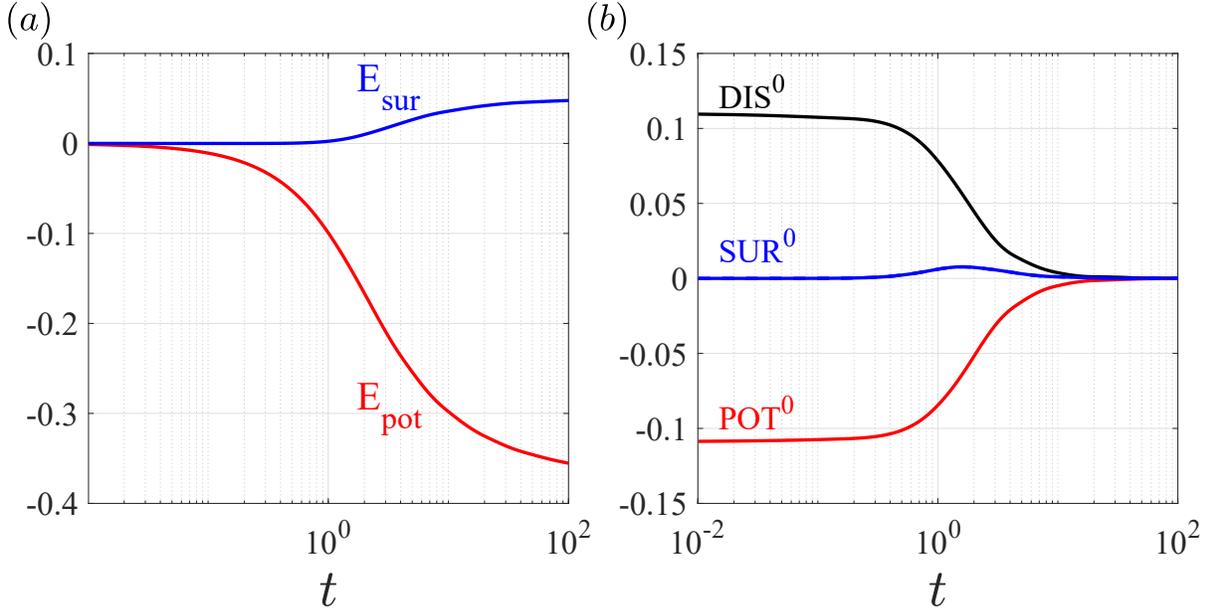}} 
	\caption{Energy analysis of the base flow presented in figure~\ref{fig:horizontalfiber-base-flow}. (a) Evolution of the surface energy $\text{E}_\text{sur}$, and potential energy $\text{E}_\text{pot}$, per unit length.
		(b) The rate of viscous dissipation $\text{DIS}^\text{0}$, surface energy $\text{SUR}^\text{0}$, and potential energy $\text{POT}^\text{0}$. 
	}
	\label{fig:horizontalfiber_baseflow_energy}
\end{figure}
%%%%%%%
The energy equation for the base flow presented in \S\ref{sec:horizontalfiber-baseflow}, computed per unit length in $z$, can be expressed as
\begin{equation} 
	\underbrace{ \iint_{\Omega_{xy}} Bo \ \delta^2 \ tr\left( { \left( \nabla {\bf u}^0 + (\nabla {\bf u}^0)^T \right) \cdot \nabla {\bf u}^0 } \right) \dAb }_{\text{DIS}^0} + \underbrace{ \int_{\p \Sigma_{\mathrm{int}}} \kappa^0 {\bf n}^0 \cdot {\bf u}^0 \dsint }_{\text{SUR}^0 } + \underbrace{ \iint_{\Omega_{xy}} -Bo \ {u}_y^0 \ \dAb}_{\text{POT}^0} = 0,
	\label{eq:horizontalfiber_energy_balance_dimless_baseflow}
\end{equation}
%%%%%%%%%%
In the case of the base flow, the dimensionless bulk potential energy per unit axial length (made dimensionless by $\gamma R^2$), evaluated with respect to the initial state, can be expressed as
\begin{equation} 
	\text{E}_\text{pot} = \iint_{\Omega_{xy}} Bo \ y \ \dAb,
	\label{eq:horizontalfiber_energy__dimlesspot}
\end{equation}
%%%%%%%%
and the dimensionless surface energy per unit axial length evaluated with respect to the initial state can be expressed as
\begin{equation} 
	\text{E}_\text{sur} = \text{SUR}^0 \big|_t - \text{SUR}^0 \big|_0 = \int_{\p \Sigma_{\mathrm{int}}} \dsint \bigg|_t -2 \pi.
	\label{eq:horizontalfiber_energy__dimlesssur}
\end{equation}
Figure~\ref{fig:horizontalfiber_baseflow_energy}(a) shows the temporal evolution of the potential and surface energies until reaching a quasi-static pendant drop for $\{\delta,Bo\}=\{0.2,0.4\}$ whose base flow is presented in figure~\ref{fig:horizontalfiber-base-flow}. As the liquid is pulled off the cylinder, the bulk potential energy is released. Being partially stored as surface energy, the potential energy allows the interface to deform. This energy transfer slows down and saturates later when the pendant curtain stagnates. The rates at which the energy is transferred are presented in figure~\ref{fig:horizontalfiber_baseflow_energy}(b)
which demonstrates that the excess potential energy is dissipated in the bulk liquid. 
%%%%%%%%%%%%%%
%%%%%%%%%%%%%%
\subsubsection{Energy analysis of the perturbed flow} \label{sec:horizontalfiber_res_energy_LSA}
%%%%%%%%%%%%%%%
%%%%%%%%%%
\begin{figure}
	\centerline{\includegraphics[width=1\textwidth]{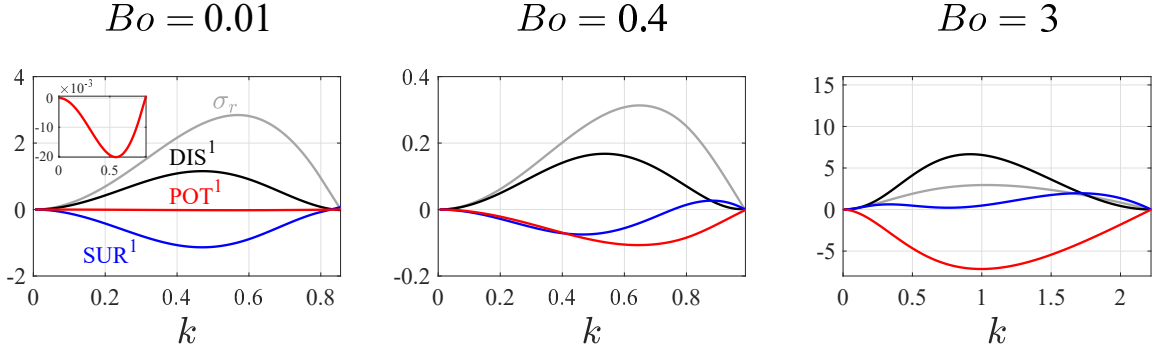}} 
	\caption{Energy analysis of the perturbations and the influence of $Bo$. The presented modes are the same as in figure~\ref{fig:horizontalfiber_LSA_overview}. Energy terms are given in (\ref{eq:horizontalfiber_energy_perturbed_final}), and the grey curve shows the dispersion curve of the mode. The inset of the left panel shows POT$^1$. All of the rate of energy terms are normalised by $\iint_{\p \Sigma_{\mathrm{int}}} | \tilde{\eta} |^2 \dsint$. 
	}
	\label{fig:horizontalfiber_LSA_energy_Bo}
\end{figure}
%%%%%%%%%
The energy equation at the scale of linear perturbations computed along one wavelength results in terms with an order of $\epsilon^2$, giving
\begin{align} \label{eq:horizontalfiber_energy_perturbed_final}
	& \left( \underbrace{ \iint_{\Omega_{xy}} Bo \ \delta^2 \ tr\left( { \left( \tilde{\nabla}  \tilde{\bf u} + (\tilde{\nabla} \tilde{\bf u})^T \right) \cdot 
    \tilde{\nabla} \tilde{\bf u}^\star } \right) \dAb  }_{\text{DIS}^1 } \right)_r  \nonumber \\
	+ & \left( \underbrace{ \int_{\p \Sigma_{\mathrm{int}}}  \tilde{\kappa} {\bf n}^0 \cdot \tilde{\bf u}^\star \dsint }_{\text{SUR}^1} + \underbrace{  \int_{\p \Sigma_{\mathrm{int}} }  \left( \tilde{\eta} \ \p_r \stau^0 \cdot {\bf n}^0 \right) \cdot \tilde{\bf u}^\star \dAb }_{\text{POT}^1  } \right)_r  =0,
\end{align}
where $^\star$ denotes the complex conjugate, ${\text{DIS}^1}$ denotes the bulk viscous dissipation rate, and ${\text{SUR}^1}$ and $\text{POT}^1$ denote the contributions of capillarity and gravity to the rate of the work done by the fluid at the perturbed interface, respectively. (For further details on the derivation of~(\ref{eq:horizontalfiber_energy_perturbed_final}) and its different terms see appendix~\ref{app:horizontalfiber_energy_pert}.) We recall that the subscript $r$ here denotes the real part of a complex number.
%%%%%%%%%%
As for the base flow, equation~(\ref{eq:horizontalfiber_energy_perturbed_final}) unravels that the work exchanged at the perturbed interface is partially dissipated in the bulk liquid, and the remainder (or deficit) is stored at (or released from) the free surface in the form of surface energy. 

%%%%%%
Figure~\ref{fig:horizontalfiber_LSA_energy_Bo} shows the results of the energy analysis on the unstable mode for $\delta=0.2$ and $Bo=\{0.01, 0.4, 3\}$. We remind the reader that the linear stability characteristics of this mode are presented in figure~\ref{fig:horizontalfiber_LSA_overview}. In the case of large surface tension to gravity ratios, typically $Bo=0.01$, where the base interface is quasi-circular, both potential and capillary mechanisms drive the instability, since their energy rates are both negative (see inset for POT$^1$).
However, the instability is dominated by capillarity, as the majority of the energy exchange to the perturbations is provided by capillarity ($|$SUR$^1| \gg$ $|$POT$^1|$). By increasing $Bo$ which results in a vertically outstretched pendant curtain, the instability becomes less favourable for the surface energy minimisation. Ultimately, exceeding a threshold for $Bo$, the instability is merely induced by the potential energy release, and capillarity acts to stabilise the flow (SUR$^1>0$ for $Bo=3$ as seen in figure~\ref{fig:horizontalfiber_LSA_energy_Bo}).
%%%%%%%%%%%%
%%%%%%%%%%%%

%%%%%%%%%%%%
\subsubsection{Asymptotic energy diagram of the static pendant curtain}  
\label{sec:horizontalfiber_res_energydiagram}
%%%%%%%%%
\begin{figure}
	\centerline{\includegraphics[width=0.6\textwidth]{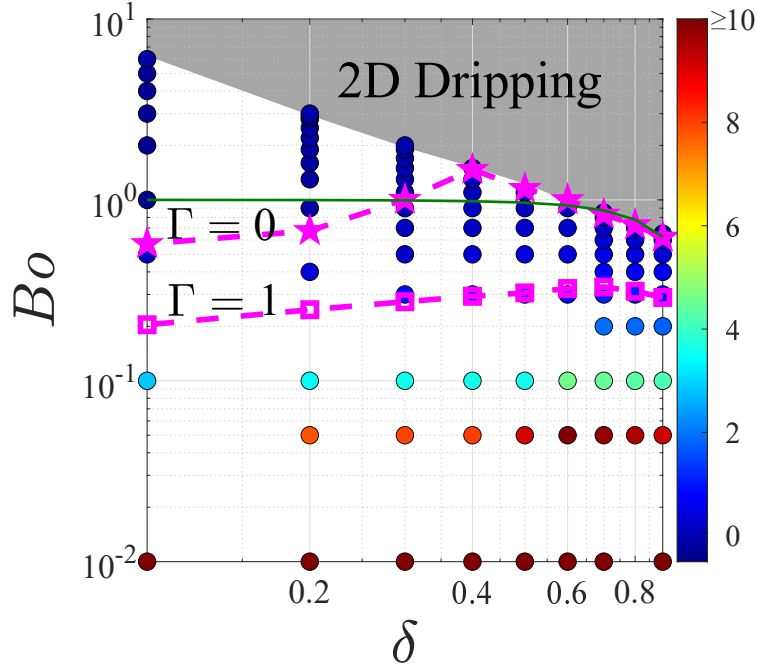}} 
	\caption{Linear stability energy diagram of the pendant curtain: the colour bar indicates the energy ratio $\Gamma$, magenta dashed lines with the pentagram and square markers show $\Gamma=0$ and $\Gamma=1$ iso-values, respectively, obtained by interpolation between linear analysis data. The sub-region above the pentagrams is purely gravity-driven Rayleigh-Taylor instability, and $Bo<0.05$ indicates capillarity-dominant instability.
    The green line shows the transition values proposed by \citet{Bruyn1997}.
	}
	\label{fig:horizontalfiber_LSA_energy_phase}
\end{figure}
%%%%%%%%%%
%%%%%%%%%%
The $\{\delta,Bo\}$ space is investigated to follow the linear stability of the pendant curtain. The pendant state is found unconditionally unstable for a single unstable mode for which the gravitational effect is always destabilising.
Here, the maximal wavenumber $\kmax$, and the contribution of the involving mechanisms are studied merely for this unstable mode. 
%%%%%
To compare the role of gravity and capillarity in the flow destabilisation, following the results presented in \S\ref{sec:horizontalfiber_res_energy_LSA}, we can define the capillary-to-potential rate of energy ratio, hereafter referred to as the {\it energy ratio}, as
\begin{equation} 
	\Gamma = \frac{ \text{SUR}^1 }{ \text{POT}^1 } \bigg| _{\kmax}.
	\label{eq:horizontalfiber_energy_ratio}
\end{equation}
%%%%%%
The energy ratio includes two pieces of information at the most linearly amplified wavenumber $\kmax$; firstly, as POT$^1<0$ always, the sign of $\Gamma$ indicates if the capillarity acts as a stabilising mechanism ($-$) or a destabilising ($+$) one.
Secondly, the magnitude of $\Gamma$ indicates if the instability is gravity-dominated ($|\Gamma| \ll 1$), capillarity-dominated ($|\Gamma| \gg 1$).
When $|\Gamma| = \mathcal{O} (1)$, both mechanisms contribute to the curtain instability. Figure~\ref{fig:horizontalfiber_LSA_energy_phase} presents the energy diagram coloured by the energy ratio. When surface tension dominates gravity, $Bo \le 0.05$, the instability is capillarity-dominated (warm colours). Although $\Gamma > 45 $ for all of the data with $Bo=0.01$, the color bar is limited to 10 for better visibility of the energy diagram. Increasing $Bo$ for a fixed $\delta$ reduces the capillary contribution.
Exceeding some threshold in $Bo$ (marked by the magenta pentagrams), capillarity becomes stabilising and the instability turns purely gravity-driven.
The sub-region of the purely gravity-driven instability narrows down by increasing $\delta$, and for $\delta \ge 0.4$ this threshold is very close and slightly inferior to the critical $Bo$ for two-dimensional curtain dripping. (The grey shaded area is the two-dimensional dripping presented in figure~\ref{fig:horizontalfiber-base-flow-Bo}(d).)

\citet{Bruyn1997} performed a leading order analysis by 
% neglecting any interactions between 
isolating
destabilising mechanisms and proposed a critical Bond number (green line in figure~\ref{fig:horizontalfiber_LSA_energy_phase}) to transition from capillary-driven instability to gravity-driven one.
This critical value corresponds to when the maximal growth rate associated with the Rayleigh-Taylor instability of a thin, flat film~\citep{Fermigier1992} dominates over that of the Rayleigh-Plateau instability of a core-annular liquid film covering a cylinder \citep{Goren1962}.
%

%%%%%%%%%
%%%%%%%%%

\subsection{Linear prediction: pattern formation and three-dimensional pinch-off} \label{sec:horizontalfiber_results_3dpinch}
%%%%%%%%%%%%%%%%%%%%%%%%%

%%%%%%%%%
\begin{figure}
	% \centerline{\includegraphics[width=0.8\textwidth]{Horizontalfiber-LSA-kmax}} 
 \centerline{\includegraphics[width=1\textwidth]{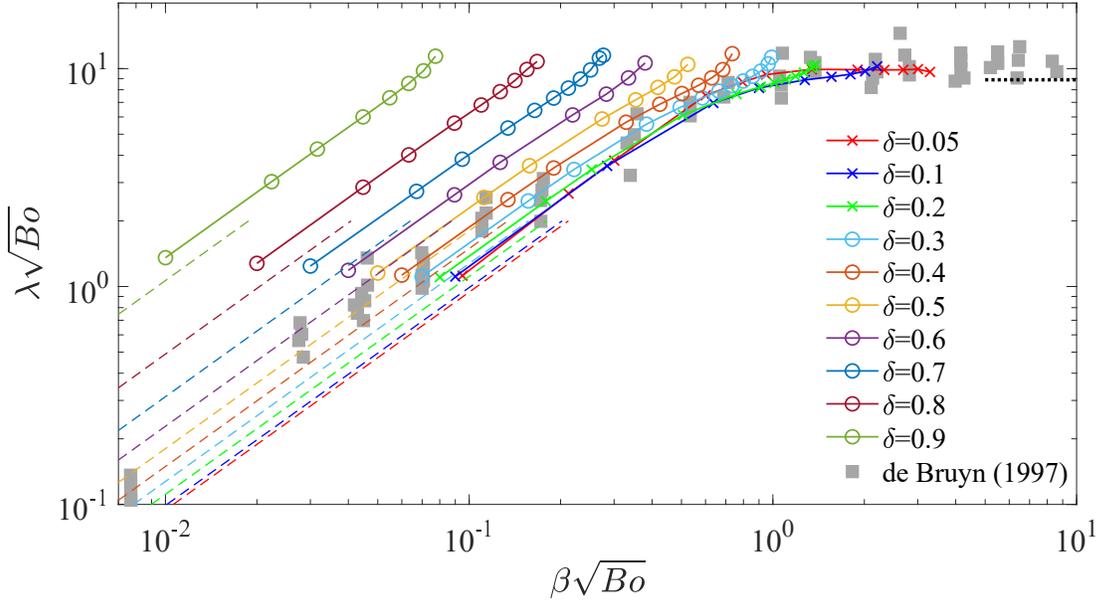}} 
	\caption{Comparison of the wavelengths observed in the experiments of~\citet{Bruyn1997} (grey squares) and the most amplified wavelength obtained from the linear analysis of the pendant curtain $\lambda_{max}$ (couloured symbols). Dashed lines show the wavelength prediction by \citet{Goren1962} for a core-annular film in absence of gravity (coulours are the same as the corresponding symbols of similar $\delta$). The black dots indicate the most unstable wavelength associated with the Rayleigh-Taylor instability of a thin liquid film under a flat plate, i.e. $\lambda_{max} \sqrt{Bo} = 2\pi \sqrt{2} \approx 8.89$ \citep{Fermigier1992}.
	}
	\label{fig:horizontalfiber_LSA_lam}
\end{figure}
%%%%%%%%%%

Figure~\ref{fig:horizontalfiber_LSA_lam} shows the comparison between the emerging wavelengths observed in the experiments of \citet{Bruyn1997} (grey squares) and the wavelengths of the most unstable perturbations, $\lambda_{max}=2\pi/\kmax$, obtained from the linear analysis of the pendant curtain (symbols) for a wide range of $\{\delta,Bo\}$. 
Those experiments were conducted by removing submerged cylinders from a liquid bath, and in the case of thin cylinders, a thin liquid film was set by paintbrush. 
As a result, for each case, the film thickness was selected naturally while detaching the covering film from the liquid bath or paintbrush, and two-dimensional curtain pinch-off was avoided in the experiments.

The axes of the figure are inspired by the analytical solution obtained by \citet{Goren1962} for a core-annular liquid film covering a cylinder in absence of gravity, represented as dashed lines. 
In the surface tension-driven regime, i.e. small Bond numbers, $\lambda_{max} \sqrt{Bo} = m(\delta) \beta \sqrt{Bo}$. 
Although this scaling is similar for a curtain and a core-annular film, the values of $m(\delta)$ are higher for a pendant curtain ($1.27 \pm 0.06$ times higher for all studied thicknesses). This difference is possibly due to the azimuthal asymmetry of the liquid bulk with respect to the solid wall, and the interface pinning at the top of the cylinder.
The prediction of our linear analysis is in agreement with the experiments of \citet{Bruyn1997} who calculated an average thickness of $\delta \approx 0.5$ for their smallest cylinders in the capillary regime (corresponding to $\beta \sqrt{Bo}<0.1$ ) and thin film at large Bond numbers by assuming that the final pendant drops were perfect spheres.

When gravity becomes prominent by increasing $Bo$, $\lambda_{max}$ deviates from the aforementioned scaling.
This deviation is less notable for $\delta > 0.5$ as the curtain experiences a two-dimensional pinch-off before reaching a quasi-stationary state.
For thinner films ($\delta<0.3$), however, the deviation from the capillary scaling is more significant at larger Bond numbers. For $\delta \le 0.1$, $\lambda_{max} \sqrt{Bo}$ seems to saturate to some values slightly larger than that of the fastest mode associated with the Rayleigh-Taylor instability of a thin film under a flat substrate, i.e. $8.89$ (black dots).
This saturation is also in accordance with the experiments of \citet{Bruyn1997},

%%%%%%%%%
\begin{figure}
	\centerline{\includegraphics[width=1\textwidth]{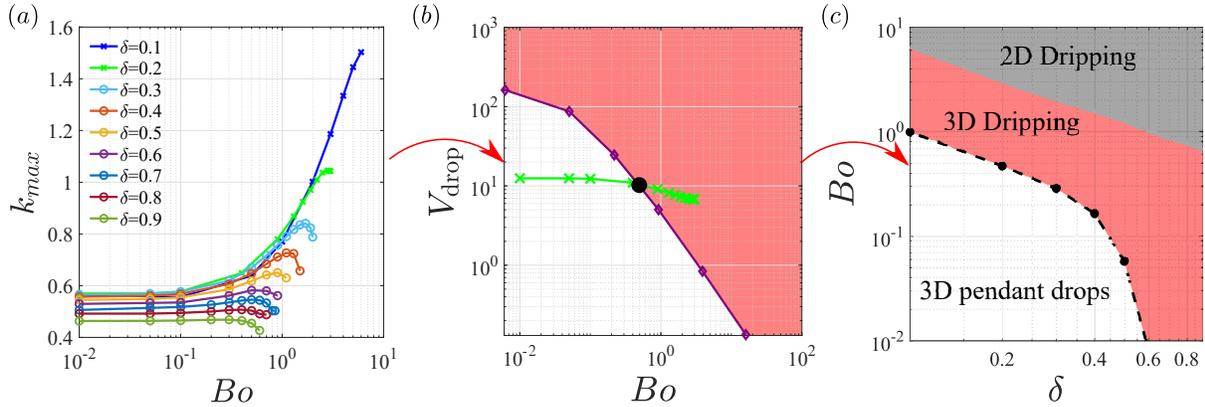}} 
	\caption{(a) $\kmax$ vs $Bo$ for different values of $\delta$ obtained from the linear analysis of the pendant curtain.
		(b) An exemplary evaluation of the drop pinching criteria for $\delta=0.2$: diamonds represent the maximum plausible volume of a static three-dimensional droplet~\citep{Weidner1997}, crosses represent the volume of a droplet based on the most linearly amplified wavenumber, $V_\text{drop}= 2 \pi^2 (1-\beta^2)/\kmax$, solid lines here show interpolation between discrete data, and the black circle indicates the critical value for pinching; the pink region highlights the drip parameters, and the white region highlight the pendant state. 
		(c) Regime diagram based on the encapsulated volume within the most linearly amplified perturbation, following the same sample calculation as in panel (b): the grey region shows the two-dimensional curtain dripping in absence of perturbations; the circles, pink and white regions indicate the post-instability state, the same as in panel (b); the black dashed line shows the logarithmic interpolation between critical values. 
	}
	\label{fig:horizontalfiber_LSA_3d}
\end{figure}
%%%%%%%%%%

%%%%%%
Further growth of linearly unstable modes forms a single array of drops, either as pearls wrapping around the cylinder, or pendant droplets underneath the cylinder which may or may not pinch off ultimately. 
The approximate volume of the biggest, single, static three-dimensional droplet that can suspend on/under a cylinder was calculated previously by \citet{Weidner1997} via non-linear simulations on thin films.
Their study evidenced that even though the amplification rate of perturbation varies with time due to non-linear effects, the fundamental wavelength of the perturbed interface does not change significantly.
They also noticed that in the case of a small Bond number, the final size of the pearls was comparable to that predicted by the linear theory for a coating film on a fibre in the absence of gravity.
Inspired by this observation, our linear analysis suggests a tentative prediction of the final state of the pendant curtain after instability, deduced from $\{\delta, Bo\}$ and $\kmax$ as follows.

Figure~\ref{fig:horizontalfiber_LSA_3d}(a) exhibits $\kmax$ predicted by the linear analysis as a function of mean film thickness and Bond number.
Mass conservation implies that the volume contained within the most linearly amplified wavelength is given by $V_\text{drop} =  \pi (1-\beta^2) (2\pi/\kmax)$.
Comparing this value with the one calculated by \citet{Weidner1997} can indicate whether or not the patterns emerging from a pendant curtain ultimately drip. 
An exemplary investigation for $\delta=0.2$ is shown in figure~\ref{fig:horizontalfiber_LSA_3d}(b), where the green crosses show the droplet volumes predicted for $\delta=0.2$ as a function of Bond number. The purple diamonds mark the biggest hanging drop volume adapted from \citet{Weidner1997}, over which pinch-off takes place. The intersection of these two curves indicates the dripping threshold.
Figure~\ref{fig:horizontalfiber_LSA_3d}(c) sketches a tentative regime diagram obtained by following a similar calculation for a wide range of $\{\delta,Bo\}$ pairs:
the grey shade indicates the parameters' range for which a two-dimensional rupture occurs, the same as presented in figure~\ref{fig:horizontalfiber-base-flow-Bo}(d). 
The pink designates the region where the most linearly amplified mode results in a volume larger than what surface tension can withstand, thus suggesting a three-dimensional pinch-off leading to dripping.
For parameter combinations in the white region, however, the pendant curtain transforms into an array of static pearls/pendant drops. 
This diagram demonstrates that the parameters in the vicinity of the two-dimensional dripping result in a three-dimensional rupture, irrespective of $\delta$. Furthermore, a thick film of $\delta>0.6$ is expected to always pinch off even when surface tension largely dominates gravity, i.e. at small $Bo$. 
%

%%%%%
We recall that the regime diagram presented in figure~\ref{fig:horizontalfiber_LSA_3d}(c) is attained from the asymptotic analysis of the quasi-static pendant curtain.
In other words, the aforementioned linear analysis does not account for non-linear effects like the possible formation of satellite pearls, or for amplification of the perturbations before reaching a two-dimensional pendant equilibrium, i.e. for the possible transient growth in the system. 
% **************\\
% **************\\
The formation of satellite pearls lowers the liquid volume encapsulated in bigger cells and therefore should reduce the risk of dripping. Moreover, as a piece of evidence, the non-linear simulations of~\citet{Weidner1997} did not report for any of their simulations the appearance of any measurable interface disturbances before reaching a pendant state. 
Later simulations of~\citet{Weidner2013} in presence of surfactants led to a similar observation.
In both studies, an initial low-level white noise of dimensionless amplitude $10^{-6}$ was applied at the interface of a liquid column at rest and concentric with the solid cylinder.
In this study, we performed a rigorous transient growth analysis toward comprehension of the short-term perturbations' amplification that confirms the wavelength selection predicted by the quasi-stationary pendant curtain analysis. (See appendix \ref{app:tansient-growth} for details about the transient growth analysis and the corresponding results.)
Therefore the present analysis should remain insightful as a conservative prediction.
Yet, it remains interesting for future studies to investigate the dynamics of the thinning liquid bridge between the two growing pearls, and possible interactions which may result in the coalescence of adjacent pearls and potentially cause secondary dripping events.
% **************\\
% **************\\

\section{Summary and conclusion} \label{sec:horizontalfiber_conclusion}
%%%%%%%%%%%

In this work, we studied the gravity-driven flow of a viscous liquid film coating the outer wall of a horizontal cylinder in the inertialess regime.
A numerical solution was first computed for the temporal evolution of an axially invariant base flow, starting from rest and of a uniform thickness.
The base flow exhibits a rapid liquid pull-off. For a fixed mean thickness, but depending on the Bond number, two trends are observed at large times: either the draining film reaches a quasi-static axially invariant pendant curtain (at small Bond numbers), or the liquid pull-off continues and results in a two-dimensional pinch-off under the cylinder (at large Bond numbers).
While at small Bond numbers, surface tension sustains a quasi-circular interface shape, increasing the Bond number results in further deformation of the interface, stretching out vertically underneath the cylinder.
A similar effect of the Bond number on the deformation of the interface was observed in~\citet{Eghbali2022in} for a liquid film coating the inside of a tube. However, inside a tube, the surrounding solid wall prohibits any two-dimensional rupture.
The critical two-dimensional pinch-off Bond number is found to scale as the inverse of the mean film thickness around the cylinder for all investigated values of thickness.
This scaling and its numerically obtained prefactor agree firmly with the critical value obtained by a scaling analysis that equates the weight of the whole liquid bulk with the surface tension in the thick-film limit. 
The scaling analysis however fails to give an accurate prediction for a thin film, as the dripping occurs following an interface necking at some distance below the cylinder. As a result, the surface tension fails to retain a fraction of the liquid bulk which accelerates below the neck.

Next, the stability of the quasi-stationary pendant curtain was investigated through a linear stability
analysis.
The curtain is found unconditionally linearly unstable. The sole unstable mode features a left/right symmetry, strong interface modulations at the bottom, and an immobile interface at the top of the cylinder for all parameter ranges.
A similar top/bottom asymmetry in the unstable mode was found in~\citet{Eghbali2022in}. In both flows, the small film thickness in the vicinity of a solid wall at the top of the interface forbids the interface from being perturbed.
The unstable eigenmode varies strongly with the Bond number;
at small Bond numbers, where surface tension dominates over gravity, the instability causes a flow reversal towards the top of the cylinder, resembling the capillary-driven Rayleigh-Plateau instability.
By increasing the Bond number, the flow reversal weakens and ultimately vanishes, and the instability promotes the formation of vertical undulations under the cylinder, resembling the Rayleigh-Taylor instability.
This observation is in accordance with the non-linear simulations of~\citet{Weidner1997,Weidner2013} in the thin film limit. 

A parametric study was then conducted to cover the space of the dimensionless parameters $\{Bo, \delta \}$, along with the energy analysis of the base flow and unstable mode.
The energy analysis helps interpreting the formation of a pendant curtain as a process to minimise the gravitational potential energy of the bulk flow, subject to a surface energy barrier.
It furthermore demonstrates that gravity is unconditionally destabilising, whereas the role of surface tension varies with the Bond number. At small Bond numbers, surface tension is destabilising and dominates the instability. The eigenflow reduces the surface energy, confirming the characteristics of the Rayleigh-Plateau instability.
In contrast, increasing the Bond number reduces the capillarity contribution to the instability, and exceeding a threshold, surface tension turns stabilising as the eigenmode reduces only the gravitational potential energy, in analogy with the purely gravity-driven Rayleigh-Taylor instability.
Both limits are in accordance with the description given by~\citet{Weidner1997} in the early stages of the appearance of disturbances to the flow in the thin film limit. Nevertheless, their nonlinear simulations addressed a stabilising effect of gravity at a low Bond number when the perturbations amplify beyond the linear range.

The present linear analysis also illustrates that the most linearly amplified wavenumber is selected through a compromise between surface energy and potential energy reduction, and varies as a function of the mean film thickness and Bond number, in agreement with the preexisting experimental measurements by ~\citet{Bruyn1997}. In the capillary regime, the most unstable wavelength follows a similar but suitably adapted scaling as that of a core-annular liquid film, and deviates from this scaling when gravity becomes comparable to the surface tension. 
Lastly, a regime diagram was sketched for the viscous liquid films draining down the outer wall of a horizontal cylinder, based on the asymptotic linear analysis of a quasi-static pendant curtain, postulating that the non-linearity does not affect the selection of the most amplified wavelength and all of the liquid contained within this wavelength concentrates in one drop.
This diagram proposes a tentative regime boundary as a function of the Bond number and the mean film thickness to predict if the emerging pattern leads into a three-dimensional pinch-off or forms a static array of wrapping pearls or pendant drops.
This prediction states that a pearl of large mean thickness eventually pinches even at a very low Bond number; an intuitively reasonable prediction. 

Finger~\citep{Takagi2010,Balestra2019} and pearl~\citep{Carroll1984,Brochard1990,McHale1999} formation in the viscous flow on top of a cylindrical substrate have been addressed extensively. However, one should note that the contact line plays an essential role in those cases, hence they differ in nature from the instabilities reported in our analysis.
Additionally, the critical pinch-off threshold may be affected by the dynamics of the meniscus connecting two adjacent drops and the possible formation and coalescence of satellite pearls. Therefore, a direct numerical simulation of a three-dimensional static drop may verify or improve the accuracy of our tentative droplet pinch-off boundary in the regime diagram.
Another direction for future investigations is to evaluate the flow instability when the cylinder is inclined so that a longitudinal component of gravity creates axial flow motion, forming a single rivulet under the cylinder~\citep{Aktershev2021}.

{\bf Declaration of interests} {The authors report no conflict of interest.}

% {\bf Funding.} {This work was supported by the Strategic Focus Area (SFA) Advanced Manufacturing, under the title of the Powder Focusing project.}

{\bf Authors' ORCID}

{S. Eghbali {\url{https://orcid.org/0000-0003-1216-2819}};

Simeon Djambov {\url{https://orcid.org/0000-0003-0979-1542}};

F. Gallaire {\url{https://orcid.org/0000-0002-3029-1457}}.}

\newpage
\appendix
\section{Derivation of the interface boundary conditions}\label{app:horizontalfiber_interfacebc}
In this section, the derivation of the interface boundary conditions is elaborated for the perturbed flow. These conditions are imposed at the perturbed interface, i.e. at$r=\Rb +\epsilon \eta^1$, while $\eta^1$ is already an unknown of the problem. By using the Taylor expansion, that is, projecting radially at the base interface, i.e. at $r=\Rb(\theta,t)$, any flow quantity at the perturbed interface can be readily approximated. This projection is referred to as flattening and for an arbitrary function $f(r,\theta,z,t)$ can be expressed as
\begin{equation} \label{eq:horizontalfiber-flattening}
	f|_{(r=\Rb+ \epsilon \eta^1 ,\theta,z,t )} = f|_{(r=\Rb,\theta,z,t )} + \epsilon \eta^1 \ \p_r f|_{(r=\Rb,\theta,z,t )} + \mathcal{O}(\epsilon^2) .
\end{equation}

By substituting the decomposed state vector of~(\ref{eq:horizontalfiber-Solution-decomposition}), into the interface conditions~(\ref{eq:horizontalfiber-kinematic-BC})-(\ref{eq:horizontalfiber-dynamic-BC}), then using the normal mode~(\ref{eq:horizontalfiber_eigenmode_ansatz}), and applying the aforementioned flattening, we can formulate these conditions as a set of equivalent constraints at the boundary of the base interface. The linearised form of the kinematic condition~(\ref{eq:horizontalfiber-kinematic-BC}) writes

\begin{equation} \label{eq:horizontalfiber-kinematic-BC-lin1}
	\p_t \left( \Rb+ \epsilon \eta^1 \right) + \left( {\bf u}^0 + \epsilon {\bf u}^1 \right)\cdot \nabla \left( \Rb+ \epsilon \eta^1 \right) = \left( {\bf u}^0 + \epsilon {\bf u}^1   \right) {\bf \cdot e}_r \quad \text{at } \ r=\Rb + \epsilon \eta^1,
\end{equation}
where the gradient vector in the Cylindrical coordinates can be expressed as $\nabla = \left( \p_r, 1/r \p_\theta, \p_z \right)^T$. 
Applying~(\ref{eq:horizontalfiber-flattening}) to~(\ref{eq:horizontalfiber-kinematic-BC-lin1}) and using the normal mode~(\ref{eq:horizontalfiber_eigenmode_ansatz}) readily results in~(\ref{eq:horizontalfiber_Interface_BC_linearised_kin}).

The linearised dynamic condition~(\ref{eq:horizontalfiber-dynamic-BC}) writes 
\begin{equation} \label{eq:horizontalfiber-dynamic-BC-lin1}
	\left( \underline{\underline{{\tau}}}^0 + \epsilon \underline{\underline{{\tau}}}^1 \right) \cdot \left( {\bf n}^0 + \epsilon {\bf n}^1 \right)  = -\left( \kappa^0 + \epsilon \kappa^1 \right) \left( {\bf n}^0 + \epsilon {\bf n}^1 \right)  \quad \text{at } \ r=\Rb + \epsilon \eta^1,
\end{equation}
%%%%%
Applying~(\ref{eq:horizontalfiber-flattening}) to~(\ref{eq:horizontalfiber-dynamic-BC-lin1}) 
and using the normal mode~(\ref{eq:horizontalfiber_eigenmode_ansatz}) readily results in~(\ref{eq:horizontalfiber_Interface_BC_linearised_dyn}). In order to express interface conditions in the Cartesian coordinates, the terms which are expressed in the Cylindrical coordinates should be transformed by employing the Jacobian transformations as 
\begin{align}
	{\bf e}_r &= \cos{\theta} \ {\bf e}_x + \sin{\theta} \ {\bf e}_y, & {\bf e}_{\theta} &= - \sin{\theta} \ {\bf e}_x + \cos{\theta} \ {\bf e}_y, \nonumber\\
	\p_r &= \cos{\theta} \ {\p}_x + \sin{\theta} \ {\p}_y, &\p_{\theta} &=  \frac{{\bf t}^0 \cdot \nabla_s}{{\bf t}^0 \cdot \nabla_s \theta },
\end{align}
where ${\bf t}^0$ denotes the unit tangent vector, and $\nabla_s = \nabla - {\bf n}^0 \left( {\bf n}^0 \cdot \nabla \right) $ is the tangential derivative at the base interface. Both conditions~(\ref{eq:horizontalfiber_Interface_BC_linearised_kin}) and (\ref{eq:horizontalfiber_Interface_BC_linearised_dyn}) include the normal vector and the curvature of the perturbed interface whose formulation is identical to that given in~\citet{Eghbali2022in}. For further details concerning the numerical implementation of the boundary conditions, see appendix~\ref{app:horizontalfiber_implementation_LSA}.

\section{Variational formulation of the linear stability analysis and implementation of its boundary conditions}\label{app:horizontalfiber_implementation_LSA}
%%%%%%%%%%%%%%
Implementation of the numerical scheme and development of the variational formulation associated with the governing equations presented in \S\ref{sec:horizontalfiber-Gov-eq} are elaborated in this appendix, recalling that the numerical domain is shown in figure~\ref{fig:horizontalfiber_numerical_domain_drainage_cartesian}. To develop the variational form of~(\ref{eq:horizontalfiber-eigenvalue-problem}), firstly the normal mode of~(\ref{eq:horizontalfiber_eigenmode_ansatz}) is applied to the system of equations~(\ref{eq:horizontalfiber-linearised-Incompressibility})-(\ref{eq:horizontalfiber_Interface_BC_linearised_kin}). Then it is internally multiplied by the vector of the test functions $\psi=[\psi_{p},\psi_{{\bf u}},\psi_{\eta}]$, where $\psi_{{\bf u}}=[ \psi_{u_x},\psi_{u_y},\psi_{u_z}]$. The resulting scalar product is integrated at $\Omega_{xy}$, which in the linear order writes
%
%%%%%%%%%%%
\begin{align} 
	& \biggl\{  \iint_{\Omega_{xy}} \psi_{p}^\star \left( \tilde{\nabla} \cdot \tilde{{\bf u}} \right) \ \mathrm{d} A_{\Omega_{xy}}  \label{eq:horizontalfiber_weak3_LSA_1}\\
	+& \iint_{\Omega_{xy}} \psi_{{\bf u}}^\star \cdot  \left( \left( \frac{Bo}{Oh} \right)^2 \delta^4 \sigma \tilde{\bf u}   \right) \ \mathrm{d} A_{\Omega_{xy}}  \\
	+& \iint_{\Omega_{xy}} \psi_{{\bf u}}^\star \cdot  \left( \left( \frac{Bo}{Oh} \right)^2 \delta^4 \left(  {\bf u}^0 \cdot \tilde{\nabla} \tilde{\bf u} + \tilde{\bf u} \cdot \nabla {\bf u}^0 \right)  \right) \ \mathrm{d} A_{\Omega_{xy}} \\ 
	+& \iint_{\Omega_{xy}} tr\left(  \tilde{\stau}^T  \cdot \left( \tilde{\nabla} \psi_{{\bf u}} \right)^\star \right) \ \mathrm{d} A_{\Omega_{xy}}   \label{eq:horizontalfiber_weak3_LSA_3}\\
	+& \int_{ \p \Sigma_\mathrm{int} }  \left(   \tilde{\stau}^0 \cdot  \tilde{\bf {n}} + \tilde{\eta}  \, \p_r \stau^0 \cdot  {\bf{n}}^0 + \left( \kappa_0 \tilde{\bf n} + \tilde{\kappa} {\bf{n}}^0 \right) \right) \cdot \psi_{{\bf u}}^\star \ \mathrm{d}s  \label{eq:horizontalfiber_weak3_LSA_4}\\
	+&  \int_{ \p \Sigma_\mathrm{int} } \psi_{\eta}^\star \left( \sigma  \tilde{\eta}  \right) \ \mathrm{d}s 	\\
	+&  \int_{ \p \Sigma_\mathrm{int} } \psi_{\eta}^\star \left( 	\left(  -\p_r u^0_r + \frac{\p_r u^0_\theta \ \p_\theta \Rb}{\Rb} - \frac{u^0_\theta \ \p_\theta \Rb }{\left(\Rb\right)^2} \right) \tilde{\eta} + \frac{u^0_\theta  }{\Rb} \p_\theta \tilde{\eta}  + \ \frac{\p_\theta \Rb}{\Rb}  \tilde{u}_{\theta} - {\tilde{u}_r } \right)  \ \mathrm{d}s \biggr\}  \\
	+& \ \text{c.c.} = 0.	\label{eq:horizontalfiber_weak3_LSA_5}
\end{align}
%%%%%%%%%%%%%%
It should be noted that in a complex system, the applied scalar product is Hermitian, defined as $\left\langle {\bf a , b} \right\rangle = {\bf a^\star \cdot b}$ where the superscript $\star$ denotes the complex conjugate.
This variational equation can be readily implemented and solved in \comsol. It is sufficient to solve the first part (in $\{ \}$) and the c.c. is known consequently. Step-by-step derivation of this equation and its matrix representation is similar to that presented in~\citet{Eghbali2022in}.

\section{Validation of the numerical model}\label{app:horizontalfiber_validation}

\begin{figure}
	\centerline{\includegraphics[width=1\textwidth]{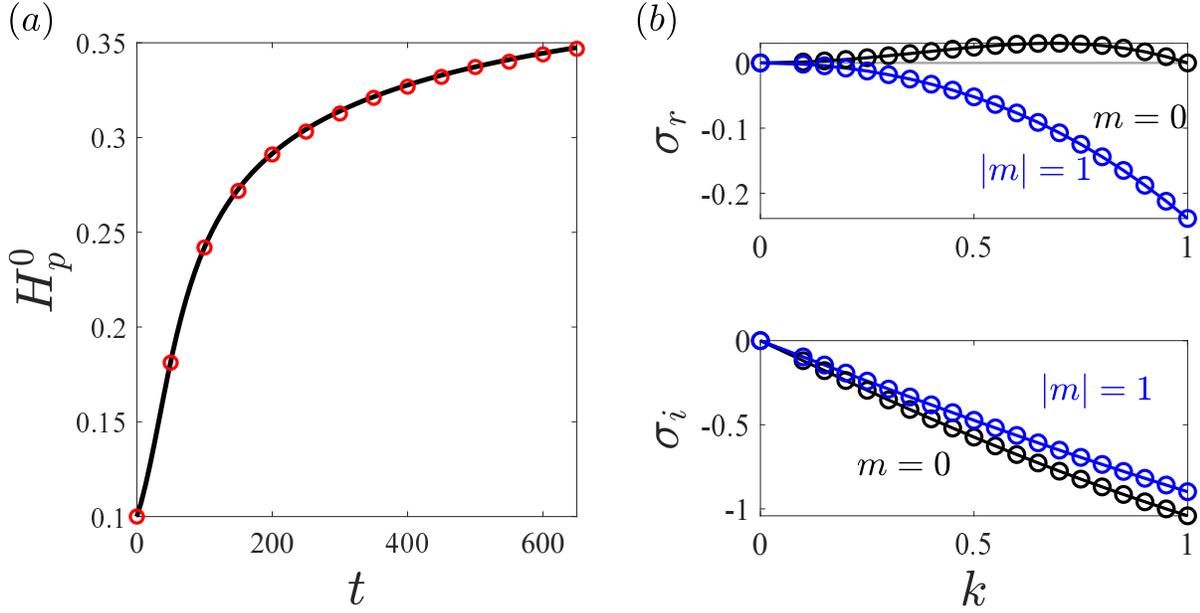}} 
	\caption{Numerical model validation; (a) base flow: temporal evolution of the south pole thickness; solid line presents the solution from the present numerical study, and the red circles present the solution obtained by~\citet{Weidner1997}; $Oh \rightarrow \infty, Bo = 1.21, \beta=10/9$; (b) linear stability analysis: dispersion curves of the two least stable modes associated with the gravity-driven viscous film flow down a centered cylinder, namely $|m|=\{0,1\}$. The continuous lines present the analytical solution obtained from the Stokes equations, and the circles represent the results from the present numerical model;~\citet{Craster2006} considered a similar perturbation as in equation~\ref{eq:horizontalfiber-Solution-decomposition} with the normal mode of $\exp[\sigma t + \mathrm{i}kz + \mathrm{i}m\theta]$, a typical choice for the axisymmetric configurations. Note that the present model, in the Cartesian coordinates, does not expand in the azimuthal wavenumber $m$; $Oh \rightarrow \infty, Bo=1, \delta=0.4$.}
	\label{fig:horizontalfiber-model-validation}
\end{figure}
The present numerical scheme is validated hereafter. Several measures are taken to ensure the correspondence of the model, based on the asymptotic limits and analytical solutions if any.
%%%%%%%%%%%%%%%%%%%%%%%%%%%%
%%%%%%%%%%%%%%%%%%%%%%%%%%%%
\subsection{Base flow model}\label{app:horizontalfiber_validation_BF}
%%%%%%%%%%%%%%%%%%%%%%%%%%%%
The present base flow model is validated with~\citet{Weidner1997} who employed the lubrication approximation~\citep{Oron1997} to simulate the non-linear gravity-driven evolution of a thin film around a solid horizontal cylinder. Figure~\ref{fig:horizontalfiber-model-validation}(a) shows the temporal evolution at the south pole of the cylinder for $\{Oh \rightarrow \infty, Bo = 1.21, \beta=10/9 \}$. The present model results in a solution of the base flow in firm agreement with the solution of~\citet{Weidner1997}. \\
%%%%%%%%%%%%%%%%%%%% \\
%%%%%%%%%%%%%%%%%%%% \\

\subsection{Linear stability analysis model}\label{app:horizontalfiber_validation_LSA}
%%%%%%%%%%%%%%%%%%%
The present linear stability model is validated with the analytical solutions that~\citet{Craster2006} presented for the gravity-driven coating flow down a vertical centered cylinder (where gravity points in $z$ direction in figure~\ref{fig:horizontalfiber_schematic}). The corresponding base flow is parallel and can be expressed in cylindrical coordinates as  
%%%%%%%
\begin{equation} \label{eq:Horizontalfiber-Nusselt-flow}
	u^0_z= \frac{\delta^{-2}}{2}  \left( \ln{\frac{r}{\beta}} - \frac{ r^2 - \beta^2}{2}  \right), \quad
	p^0 = 1, \quad
	\Rb = 1.
\end{equation}
%%%%%%%
For the linear stability analysis,~\citet{Craster2006} employed the long-wavelength approximation~\citep{Reynolds1886} and compared the results with the analytical solution, in terms of Bessel functions, obtained by solving the full Stokes equations~\citep{Goren1962}. Unlike the present study that is formulated in the Cartesian coordinates,~\citet{Craster2006} used the axisymmetry of the flow and considered a perturbation as in (\ref{eq:horizontalfiber_eigenmode_ansatz}) with the normal mode exponent of $\exp[\sigma t + \mathrm{i}kz + \mathrm{i}m\theta]$ where $m$ denotes the azimuthal wavenumber. Figure~\ref{fig:horizontalfiber-model-validation}(b) presents the agreement between the present linear stability model and the analytical solution for a thick film $\delta=0.4$. It should be noted that despite the axisymmetric nature of the validated case, this presented validation holds also for an arbitrary interface. For this aim, the geometrical symmetry in the numerical reference frame is broken by setting the origin of the coordinates system at an arbitrary location inside the liquid film, $(x,y)=(0.2,0.7)$. 
%%%%%%%%%%%%%%%%%%%%%%%
%%%%%%%%%%%%%%%%%%%%%%%%%%
  
\subsection{Grid independency}\label{app:horizontalfiber_validation_mesh}
\begin{figure}
	\centerline{\includegraphics[width=1\textwidth]{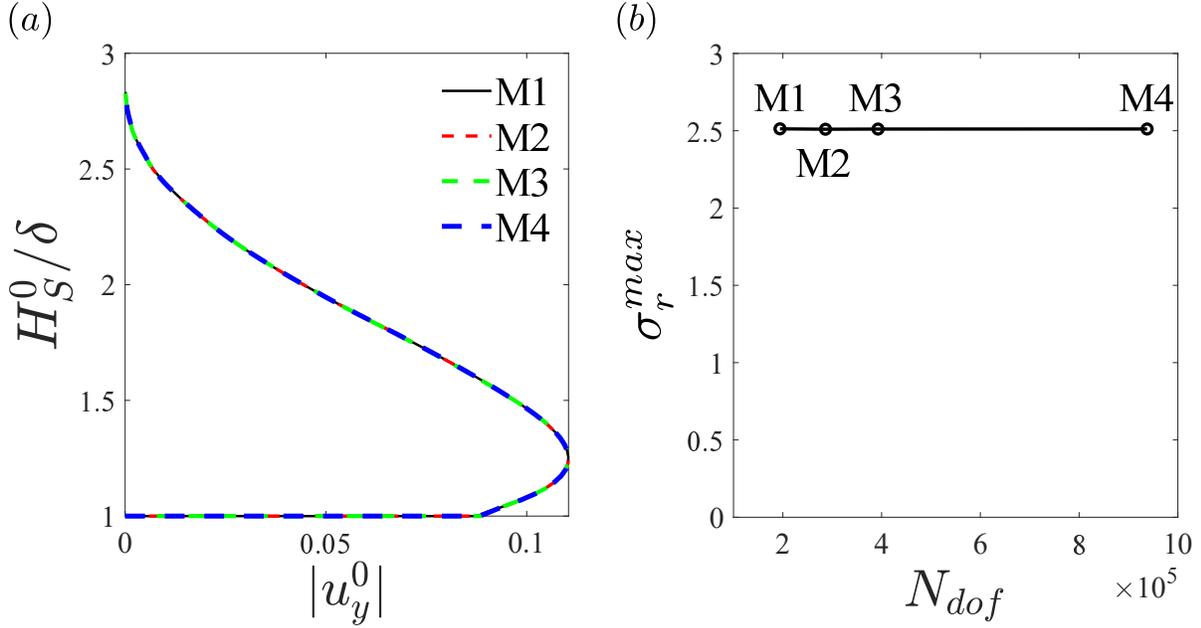}}%
	\caption{Mesh convergence proof for $Oh\rightarrow \infty$, $Bo=0.4$, $\delta=0.2$: (a) base flow; $H_S^0 / \delta$ vs $|u_y^0|$; (b) linear stability analysis; $\sigma_r^{max}$ vs $N_{dof}$. All of the results presented in this manuscript are obtained from M3.}
	\label{fig:horizontalfiber_mesh_dep}
\end{figure}

A convergence study for the base flow evolution and the linear stability of the most unstable eigenvalue is presented in figure~\ref{fig:horizontalfiber_mesh_dep}, for $\{ Oh \rightarrow \infty, Bo=0.4, \delta = 0.2 \}$. Mesh resolution is controlled by setting a prefactor multiplied by the number of divisions at the solid wall and interface boundaries. Mesh convergence is obtained for the presented grids. All of the presented results in the manuscript are obtained employing M3.
%%%%%%%%
%%%%%%%%
\section{Derivation of the energy equation}\label{app:horizontalfiber_energyderiv}
%%%%%%%%%
%%%%%%%%%

In this section, the derivation of the energy equation is elaborated. 
Dimensional form of the momentum equation~(\ref{eq:horizontalfiber-Navier-Stokes-dimless}) in the inertialess limit is given in~\citet{Eghbali2022}.
Following the same formalism, the dimensionless form of the energy equation, obtained under the scaling presented in \S\ref{sec:horizontalfiber-Gov-eq}, can be expressed as

%%%%%%%%%%%%
\begin{equation} 
	\underbrace{ \iiint_{\Omega_{xy}} Bo \ \delta^2 \ tr\left( \left( \nabla {\bf u} + (\nabla {\bf u)^T } \right) \cdot{\bf \nabla u } \right) \text{d}V}_{\text{DIS}} + \underbrace{ \iint_{\p \Sigma_\text{j}} -\left( \stau \cdot {\bf n}^0 \right) \cdot {\bf u} \ \mathrm{d} A_{\Sigma_\text{j}}}_{\text{BND} } + \underbrace{ \iiint_{\Omega_{xy}}  - Bo \ u_y \text{d}V }_{\text{POT}} = 0.
	\label{eq:horizontalfiber_energy_balance_dimless}
\end{equation}
% %%%%%%%%%%%%%
Each under-brace denotes the physical mechanism associated with the respective term, as follows:
\begin{enumerate} 
	\item DIS: the rate of viscous dissipation in the bulk fluid
	\item BND: the rate of work done by the fluid through the moving boundaries
	\item POT: the rate of change of gravitational potential energy\\
\end{enumerate}
%%%%%%%%%
%%%%%%%%%%

No-slip condition implies ${\bf u=0}$ at $\p \Sigma_{\mathrm{w}}$, thus yielding~(\ref{eq:horizontalfiber_energy_balance_dimless_simp}).
%%%%%%%%%
%%%%%%%%%

\subsection{Energy equation for the perturbed flow}\label{app:horizontalfiber_energy_pert}
%%%%%%%%%%
The energy equation for the perturbed flow is obtained by substituting the perturbed state vector~(\ref{eq:horizontalfiber-Solution-decomposition}) with the normal mode~(\ref{eq:horizontalfiber_eigenmode_ansatz}), into~(\ref{eq:horizontalfiber_energy_balance_dimless_simp}) and integrating it over one wavelength $\Delta z=\lambda ={2 \pi}/{k}$. The resulting integral is of the order $\epsilon^2$ and determines the energy equation for the linear perturbations which implies

\begin{equation}
	\frac{2\pi}{k} \exp({2 \sigma_r t})	\left[ \underbrace{ \iint_{\Omega_{xy}} Bo \ \delta^2 \ tr\left( { \left( \tilde{\nabla} \tilde{\bf u} + (\tilde{\nabla} \tilde{\bf u})^T \right) \cdot \tilde{\nabla} \tilde{\bf u}^\star } \right)  \dAb}_{\text{DIS}^1 } + \underbrace{  \int_{\p \Sigma_{\mathrm{int}} }  -\left( \stau \cdot {\bf n}^0 \right) \cdot \tilde{\bf u}^\star \ \dsint}_{\text{BND}^1 } \right] + \text{c.c.} = 0,
	\label{eq:horizontalfiber_energy_perturbed_1}
\end{equation}
%%%%%
We remind that the normal mode~(\ref{eq:horizontalfiber_eigenmode_ansatz}) is complex, hence the integral of terms in $\epsilon^1$ order vanish due to the periodicity of the perturbations over $\lambda$. As $({2\pi}/{k}) \exp({2 \sigma_r t}) > 0$, it can be factorised and simplified. We hereafter only focus on the real part of~(\ref{eq:horizontalfiber_energy_perturbed_1}) which writes
\begin{equation}
	\left( \text{DIS}^1 + \text{BND}^1 \right)_r  = 0.
	\label{eq:horizontalfiber_energy_perturbed_2}
\end{equation}
%%%%%%%%%%%%%5
Let us recall~(\ref{eq:horizontalfiber_Interface_BC_linearised_dyn}), and that for a quasi-static pendant drop where ${\bf u}^0 \approx \bf 0$, $\stau^0 = -p^0 \tens{I}$. Except for the thin film covering the upper side of the cylinder, where $\tilde{\eta} \approx 0$ and pressure follows the lubrication pressure, bulk pressure is hydrostatic, $p^0 \approx -Bo \ y$. Thus, BND$^1$ can be decomposed as 
\begin{equation}
	\text{BND}^1=\text{SUR}^1+\text{POT}^1,
	\label{eq:horizontalfiber_energy_perturbed_BND}
\end{equation}
%%%%%%%%%
where SUR$^1$ denotes the capillary contribution to the rate of work at the perturbed interface, and POT$^1$ denotes the rate at which the fluid works against the hydrostatic pressure to perturb the interface. These terms can be expressed as

\begin{equation} \label{eq:horizontalfiber_energy_perturbed_SUR1_app}
	\text{SUR}^1 = \int_{\p \Sigma_{\mathrm{int}}}  \tilde{\kappa} {\bf n}^0 \cdot \tilde{\bf u}^\star \dsint,
\end{equation}
%%%%%%%%
\begin{equation} \label{eq:horizontalfiber_energy_perturbed_POT1}
	\text{POT}^1 = \int_{\p \Sigma_{\mathrm{int}} }  \left( \tilde{\eta} \ \p_r \stau^0 \cdot {\bf n}^0 \right) \cdot \tilde{\bf u}^\star \dsint,
\end{equation}
%%%%%%%%%%
thus giving~(\ref{eq:horizontalfiber_energy_perturbed_final}).
%%%%%%%%%%%%%%%%%%%%%%%%

\section{Transient growth analysis}\label{app:tansient-growth}
%%%%%%%%%%%
\begin{figure}
	\centerline{\includegraphics[width=0.95\textwidth]{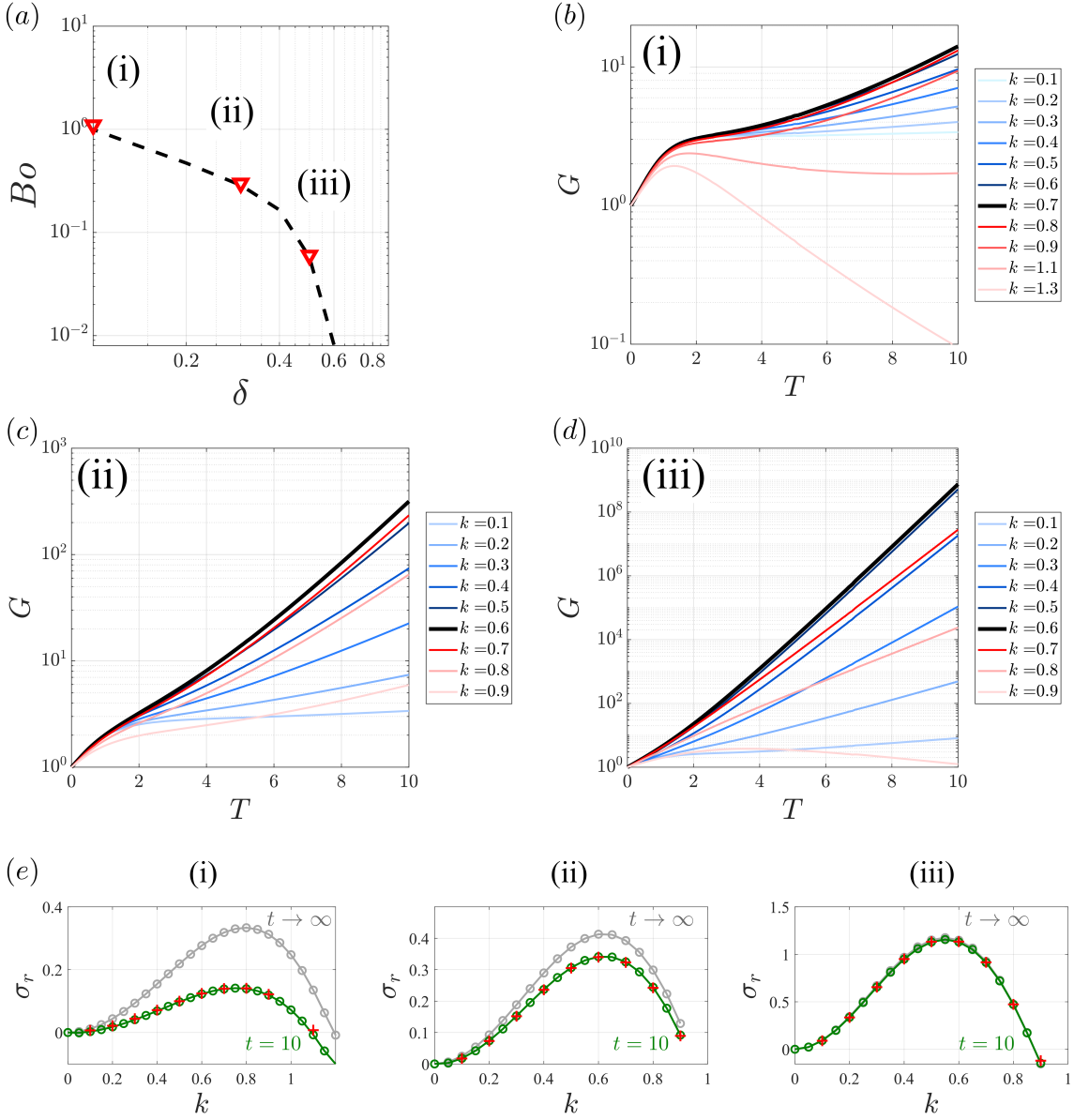}} 
	\caption{Transient growth analysis: (a) Three pinching pairs of $\{Bo,\delta\}$ analysed in the vicinity of the critical pinch-off separatrix (dashed line):
		(b) Point (i): $\{Bo,\delta\}=\{1.1,0.1\}$;
		(c) Point (ii): $\{Bo,\delta\}=\{0.3,0.3\}$
		(d) Point (iii): $\{Bo,\delta\}=\{0.06,0.5\}$.
		Panels (b-d) present the optimal transient gain $G$, as a function of the temporal horizon $T$, for different wave numbers $k$. The thick black line represents the transient gain of the most amplified wavenumber at $T=10$.
        (e) Dispersion curves of the same points obtained by the linear stability analysis of the quasi-stationary curtain, $t\rightarrow \infty$ (grey circles), linear stability analysis of the frozen frame at $t=10$ (green circles), and from the slope of the optimal gain at $t=10$ (red crosses).
  The exponential evolution of the asymptotically predicted most unstable wave number $\kmax$, is shown in a black dashed line, and 
	}
	\label{fig:horizontalfiber_tg}
\end{figure}
%%%%%%%%%%%
As the base flow presented in \S\ref{sec:horizontalfiber-baseflow} evolves temporally, we perform a \textit{transient growth analysis} to study the evolution of the perturbations from the initial state until the formation of the pendant curtain. In contrast with the linear stability analysis, the transient growth analysis accounts for the temporal dependency of both the base flow and perturbations. 

% We follow the procedure of~\citet{Eghbali2022in};
% ~\citet{Eghbali2021internal}:
% the perturbation is now written without imposing its exponential temporal evolution
Exponential growth is no longer imposed by the perturbation ansatz
%%%%%%%%%%%%%%%%%%%%%%%%%%%%%%%%%%%%%%%%%%%%%%
%%%%%%%%%%%%%%%%%%%%%%%%%%%%%%%%%%%%%%%%%%%%%%
\begin{equation} 
	{\bf q}^1=\hq(t,x,y) \ \mathrm{exp} \left[\mathrm{i}kz \right] + \mathrm{c.c.},
	\label{eq:horizontalfiber_tg_ansatz}
\end{equation} 
%%%%%%%%%%%%%%%%%%%%%%%%%%%%%%%%%%%%%%%%%%%%%%
%%%%%%%%%%%%%%%%%%%%%%%%%%%%%%%%%%%%%%%%%%%%%%
% which is instead a straightforward extension of (\ref{eq:horizontalfiber-eigenvalue-problem}):
the evolution of which is instead a straightforward extension of (\ref{eq:horizontalfiber-eigenvalue-problem}) as
%%%%%%%%%%%%%%%%%%%%%%%%%%%%%%%%%%%%%%%%%%%%%%
%%%%%%%%%%%%%%%%%%%%%%%%%%%%%%%%%%%%%%%%%%%%%%
\begin{equation} \label{eq:horizontalfiber-tg-ode}
	\tens{L}(t) \hq + \text{c.c.} = \tens{B} {\partial_t \hq} + \text{c.c.},
\end{equation}
%%%%%%%%%%%%%%%%%%%%%%%%%%%%%%%%%%%%%%%%%%%%%%
%%%%%%%%%%%%%%%%%%%%%%%%%%%%%%%%%%%%%%%%%%%%%%
recalling that $\tens{L}(t)$ is also parametrised by $k$. We seek an initial perturbation of the interface $\hq(0,x,y)=[{\bf 0},0,\bn(0)]^T$, that is the most amplified by (\ref{eq:horizontalfiber-tg-ode}) after a time $t=T$, where $T$ is named the \textit{temporal horizon}, with respect to some objective function, for which we chose the associated interfacial energy density per unit axial wavelength
%  To this aim, we pursue the methodology proposed by~\citet{DelGuercio2014}, and take advantage of the fact that the base flow interface is axisymmetric at $t=0$ (see figure \ref{fig:horizontalfiber-base-flow}) to expand $\bn(0)$ as Fourier modes in $\theta$ as
%%%%%%%%%%%%%%%%%%%%%%%%%%%%%%%%%%%%%%%%%%%%%%
%%%%%%%%%%%%%%%%%%%%%%%%%%%%%%%%%%%%%%%%%%%%%%
% \begin{equation} \label{eq:horizontal-n0}
% 	\bn(0) = \sum_{m=-N}^{N} \al_m e^{\mathrm{i}m\theta} = a_0 + \sum_{m=1}^{N} [a_m \cos(m\theta) + b_m \sin(m\theta)],
% \end{equation}
%%%%%%%%%%%%%%%%%%%%%%%%%%%%%%%%%%%%%%%%%%%%%%
%%%%%%%%%%%%%%%%%%%%%%%%%%%%%%%%%%%%%%%%%%%%%%
% with $\alpha_{-m} = \al_m^{\star}$, $a_0=\alpha_0$, and for $m\geq 1$: $a_m = \al_m+\al_m^{\star}$ and $b_m = \text{i}(\al_m-\al_m^{\star})$. Note that $a_m$, $b_m$ (where $m=0,1,2,...$) and $\bn(0)$ are real-valued.

% We then define $\bn_m(t)$ as the evolved state at time $t$ of the specific initial condition $\bn_m(0)=e^{\mathrm{i} m \theta}$ for the interface. Thanks to the linearity of the evolution equation (\ref{eq:horizontalfiber-tg-ode}), the interface shape at $t=T$ simply reads
%%%%%%%%%%%%%%%%%%%%%%%%%%%%%%%%%%%%%%%%%%%%%%
%%%%%%%%%%%%%%%%%%%%%%%%%%%%%%%%%%%%%%%%%%%%%%
% \begin{equation} 
% 	\bn(T) = \sum_{m=-N}^{N} \al_m \bn_m(T).
% \end{equation}
%%%%%%%%%%%%%%%%%%%%%%%%%%%%%%%%%%%%%%%%%%%%%%
%%%%%%%%%%%%%%%%%%%%%%%%%%%%%%%%%%%%%%%%%%%%%%
% The associated interfacial energy density per spanwise wavelength is proportional to
%%%%%%%%%%%%%%%%%%%%%%%%%%%%%%%%%%%%%%%%%%%%%%
%%%%%%%%%%%%%%%%%%%%%%%%%%%%%%%%%%%%%%%%%%%%%%
\begin{equation}
	e(T) = \frac{k}{2 \pi} \int_{\p \Sigma_{\mathrm{int}} } \int_{0}^{2\pi/k} |\bn(T)e^{\mathrm{i}kz} + \text{c.c.} |^2 \mathrm{d}z \mathrm{d}s. 
 % = 2\bm{a}^T\mE(T)\bm{a}, 
\end{equation}
%%%%%%%%%%%%%%%%%%%%%%%%%%%%%%%%%%%%%%%%%%%%%%
%%%%%%%%%%%%%%%%%%%%%%%%%%%%%%%%%%%%%%%%%%%%%%
% where $\oint_{\mathcal{S}(T)}\mathrm{d}s$ represents the closed line integration along the time-dependent base state interface. 
% , $\mE(T)$ is a real-valued, symmetric, strictly positive definite $(2N+1)\times(2N+1)$ matrix, derived in appendix~\ref{app:horizontalfiber-tgmatrix}, and
% \begin{equation}
% 	\bm{a} = [a_N,a_{N-1},...,a_1,\sqrt{2}a_0,b_N,b_{N-1},...,b_1].
% \end{equation}
% 
The \textit{optimal transient gain} is thus defined as
%%%%%%%%%%%%%%%%%%%%%%%%%%%%%%%%%%%%%%%%%%%%%%
%%%%%%%%%%%%%%%%%%%%%%%%%%%%%%%%%%%%%%%%%%%%%%
\begin{equation}
	G(T) = \max_{\bn(0)}\frac{e(T)}{e(0)}. 
 % = \pi^{-1}\frac{\bm{a}^T\mE(T)\bm{a}}{\bm{a}^T\bm{a}},
	\label{eq:horizontalfiber-tgdef}
\end{equation}
%%%%%%%%%%%%%%%%%%%%%%%%%%%%%%%%%%%%%%%%%%%%%%
%%%%%%%%%%%%%%%%%%%%%%%%%%%%%%%%%%%%%%%%%%%%%%
% is simply the largest eigenvalue of $\mE(T)$ divided by $\pi$, and the associated eigenvector provides directly the Fourier mode coefficients of the optimal initial condition, where we used that $\mE(0)=\pi\tens{I}$.

Imperfect initial flow structure or experimental artifacts might project on this optimal one and might be greatly amplified, possibly triggering a non-linear regime and/or being directly comparable with the experimentally observed patterns, thus making the linear stability analysis results irrelevant.
In practice, we make use of the initial condition's axisymmetry and decompose the initial perturbation into Fourier modes. The simulation of the linear evolution of these modes allows us to construct a propagator matrix, the leading eigenvalue of which is directly the optimal transient gain, with the Fourier coefficients of the optimal initial condition as its associated eigenvector. (See~\citet{Eghbali2022in} for more details.)
%
% This is in contrast with the quasi-stationary or {\it frozen-frame} approach, whereby exponential growth is still imposed on the perturbation's evolution, but with a rate $\sigma$, parametrically dependent on time. At each separate frame, a dispersion relation can be obtained from \eqref{eq:horizontalfiber-tg-ode} on the frozen base state. As the base state saturates towards its stationary pendant curtain shape, these frozen frame dispersion relations will also converge onto the asymptotic dispersion relation.
This is in contrast with the {\it frozen frame} approach, whereby a perturbation normal mode as in \eqref{eq:horizontalfiber_eigenmode_ansatz} is still applied to each time instant of the base flow (i.e. frame) but with an evolving rate $\sigma$, parametrically dependent on time. At each separate frame, a dispersion relation can be obtained from \eqref{eq:horizontalfiber-eigenvalue-problem} on the frozen base state pertaining to this time-instant.
Such an assumption becomes more relevant when the base flow evolves significantly slower than the perturbations~\citep{Eghbali2022in}. As in this study, the base flow saturates towards its stationary pendant curtain state, it is expected, yet to be verified, that the transient evolution of the perturbations will ultimately converge onto the asymptotic dispersion relation associated with the frozen frames. 

For the transient growth analysis, the linearised conservation equations (\ref{eq:horizontalfiber-tg-ode}) are solved over the corresponding time horizon with a dimensionless time step of 0.1. Then, the resulting propagator matrices are computed and imported to Matlab$^\text{TM}$, where the optimal transient gain (\ref{eq:horizontalfiber-tgdef}) is computed (similar to \citep{Eghbali2022in}). 
For the transient growth analysis, the computation of the propagator matrices with $-5  \le m  \le 5$ over several values of $k$.

We present the results of the transient growth analysis for three pairs of parameters $\{Bo,\delta\}$, shown in figure~\ref{fig:horizontalfiber_tg}(a), in the vicinity of the three-dimensional pinch-off separatrix, obtained in \S \ref{sec:horizontalfiber_results_3dpinch}.
As it is constructed using the drop volumes, corresponding to the most unstable wavelengths, predicted by the asymptotic frozen-frame linear stability analysis of the pendant state, it is especially important to evaluate any transient effects, which might amplify potentially different wavelengths before the full saturation of the pendant curtain, thus possibly modifying the size and end-fate of the drops.
Figures~\ref{fig:horizontalfiber_tg}(b-d) present the optimal transient gains as a function of the time horizon, for different wave numbers $k$. We observe no appreciable transient growth, which is consistent with the lack of measurable interface disturbances before the saturation of the pendant curtain in the non-linear simulations of~\citet{Weidner1997}. 
In each panel, the solid black line represents the large-time asymptotically most unstable wave number predicted by linear stability analysis among the discretised values of $k$ (as given in figure \ref{fig:horizontalfiber_LSA_3d}(a)). It is seen to also correspond to
% .   coincides with that of the envelope, i.e. it is as amplified as 
the transiently most amplified wave number. 

Finally, figures~\ref{fig:horizontalfiber_tg}(e) compare the optimal transient growth rates of the different wave numbers, calculated from the slopes in the log-linear optimal transient gain plots, at the time horizon $T=10$ (crosses), to the frozen-frame dispersion relations, obtained for $t=10$ (shown in green) and for the saturated curtain $t\to\infty$ (shown in grey). This suggests that, already at $t=10$, the system's evolution appears to be well captured by the quasi-stationary stability analyses. It is reduced to the slow succession of the dispersion relations, frozen frame after frozen frame. Furthermore, since these dispersion relations seem to retain their qualitative shape, the wavelength selection does not appear to be affected by the transient nature of the drainage flow.
{We conclude that the asymptotic, quasi-stationary, modal stability analysis should be sufficient for the tentative determination of a criterion for three-dimensional pinch-off.}

% \backsection[Declaration of interests]{The authors report no conflict of interest.}

% \backsection[Author ORCID]{S. Eghbali {\url{https://orcid.org/0000-0003-1216-2819}}; \\
% 	Y-M. Ducimeti\`{e}re {\url{https://orcid.org/0000-0003-4018-5751}}; \\
% 	E. Boujo {\url{https://orcid.org/0000-0002-4448-6140}};\\
% 	F. Gallaire {\url{https://orcid.org/0000-0002-3029-1457}}.}

\bibliography{apssamp}% Produces the bibliography via BibTeX.

\providecommand{\noopsort}[1]{}\providecommand{\singleletter}[1]{#1}%
\begin{thebibliography}{49}%
\makeatletter
\providecommand \@ifxundefined [1]{%
 \@ifx{#1\undefined}
}%
\providecommand \@ifnum [1]{%
 \ifnum #1\expandafter \@firstoftwo
 \else \expandafter \@secondoftwo
 \fi
}%
\providecommand \@ifx [1]{%
 \ifx #1\expandafter \@firstoftwo
 \else \expandafter \@secondoftwo
 \fi
}%
\providecommand \natexlab [1]{#1}%
\providecommand \enquote  [1]{``#1''}%
\providecommand \bibnamefont  [1]{#1}%
\providecommand \bibfnamefont [1]{#1}%
\providecommand \citenamefont [1]{#1}%
\providecommand \href@noop [0]{\@secondoftwo}%
\providecommand \href [0]{\begingroup \@sanitize@url \@href}%
\providecommand \@href[1]{\@@startlink{#1}\@@href}%
\providecommand \@@href[1]{\endgroup#1\@@endlink}%
\providecommand \@sanitize@url [0]{\catcode `\\12\catcode `\$12\catcode
  `\&12\catcode `\#12\catcode `\^12\catcode `\_12\catcode `\%12\relax}%
\providecommand \@@startlink[1]{}%
\providecommand \@@endlink[0]{}%
\providecommand \url  [0]{\begingroup\@sanitize@url \@url }%
\providecommand \@url [1]{\endgroup\@href {#1}{\urlprefix }}%
\providecommand \urlprefix  [0]{URL }%
\providecommand \Eprint [0]{\href }%
\providecommand \doibase [0]{https://doi.org/}%
\providecommand \selectlanguage [0]{\@gobble}%
\providecommand \bibinfo  [0]{\@secondoftwo}%
\providecommand \bibfield  [0]{\@secondoftwo}%
\providecommand \translation [1]{[#1]}%
\providecommand \BibitemOpen [0]{}%
\providecommand \bibitemStop [0]{}%
\providecommand \bibitemNoStop [0]{.\EOS\space}%
\providecommand \EOS [0]{\spacefactor3000\relax}%
\providecommand \BibitemShut  [1]{\csname bibitem#1\endcsname}%
\let\auto@bib@innerbib\@empty
%</preamble>
\bibitem [{\citenamefont {de~Bruyn}(1997)}]{Bruyn1997}%
  \BibitemOpen
  \bibfield  {author} {\bibinfo {author} {\bibfnamefont {J.~R.}\ \bibnamefont
  {de~Bruyn}},\ }\bibfield  {title} {\bibinfo {title} {Crossover between
  surface tension and gravity-driven instabilities of a thin fluid layer on a
  horizontal cylinder},\ }\href@noop {} {\bibfield  {journal} {\bibinfo
  {journal} {Phys. Fluids}\ }\textbf {\bibinfo {volume} {9(6)}},\ \bibinfo
  {pages} {1599} (\bibinfo {year} {1997})}\BibitemShut {NoStop}%
\bibitem [{\citenamefont {Weidner}\ \emph {et~al.}(1997)\citenamefont
  {Weidner}, \citenamefont {Schwartz},\ and\ \citenamefont
  {Eres}}]{Weidner1997}%
  \BibitemOpen
  \bibfield  {author} {\bibinfo {author} {\bibfnamefont {D.~E.}\ \bibnamefont
  {Weidner}}, \bibinfo {author} {\bibfnamefont {L.~W.}\ \bibnamefont
  {Schwartz}},\ and\ \bibinfo {author} {\bibfnamefont {M.~H.}\ \bibnamefont
  {Eres}},\ }\bibfield  {title} {\bibinfo {title} {Simulation of coating layer
  evolution and drop formation on horizontal cylinders},\ }\href@noop {}
  {\bibfield  {journal} {\bibinfo  {journal} {J. Colloid Interface Sci.}\
  }\textbf {\bibinfo {volume} {187}},\ \bibinfo {pages} {243} (\bibinfo {year}
  {1997})}\BibitemShut {NoStop}%
\bibitem [{\citenamefont {Elettro}\ \emph {et~al.}(2016)\citenamefont
  {Elettro}, \citenamefont {Neukirch}, \citenamefont {Vollrath},\ and\
  \citenamefont {Antkowiak}}]{Elettro2016}%
  \BibitemOpen
  \bibfield  {author} {\bibinfo {author} {\bibfnamefont {H.}~\bibnamefont
  {Elettro}}, \bibinfo {author} {\bibfnamefont {S.}~\bibnamefont {Neukirch}},
  \bibinfo {author} {\bibfnamefont {F.}~\bibnamefont {Vollrath}},\ and\
  \bibinfo {author} {\bibfnamefont {A.}~\bibnamefont {Antkowiak}},\ }\bibfield
  {title} {\bibinfo {title} {In-drop capillary spooling of spider capture
  thread inspires hybrid fibers with mixed solid--liquid mechanical
  properties},\ }\href@noop {} {\bibfield  {journal} {\bibinfo  {journal}
  {Proceedings of the National Academy of Sciences}\ }\textbf {\bibinfo
  {volume} {113}},\ \bibinfo {pages} {6143} (\bibinfo {year}
  {2016})}\BibitemShut {NoStop}%
\bibitem [{\citenamefont {Herwitz}(1987)}]{Herwitz1987}%
  \BibitemOpen
  \bibfield  {author} {\bibinfo {author} {\bibfnamefont {S.~R.}\ \bibnamefont
  {Herwitz}},\ }\bibfield  {title} {\bibinfo {title} {Raindrop impact and water
  flow on the vegetative surfaces of trees and the effects on stemflow and
  throughfall generation},\ }\href@noop {} {\bibfield  {journal} {\bibinfo
  {journal} {Earth surface processes and landforms}\ }\textbf {\bibinfo
  {volume} {12}},\ \bibinfo {pages} {425} (\bibinfo {year} {1987})}\BibitemShut
  {NoStop}%
\bibitem [{\citenamefont {Qu{\'e}r{\'e}}(1999)}]{Quere1999}%
  \BibitemOpen
  \bibfield  {author} {\bibinfo {author} {\bibfnamefont {D.}~\bibnamefont
  {Qu{\'e}r{\'e}}},\ }\bibfield  {title} {\bibinfo {title} {Fluid coating on a
  fiber},\ }\href@noop {} {\bibfield  {journal} {\bibinfo  {journal} {Annu.
  Rev. Fluid Mech.}\ }\textbf {\bibinfo {volume} {31}},\ \bibinfo {pages} {347}
  (\bibinfo {year} {1999})}\BibitemShut {NoStop}%
\bibitem [{\citenamefont {Shen}\ \emph {et~al.}(2002)\citenamefont {Shen},
  \citenamefont {Gleason}, \citenamefont {McKinley},\ and\ \citenamefont
  {Stone}}]{Shen2002}%
  \BibitemOpen
  \bibfield  {author} {\bibinfo {author} {\bibfnamefont {A.~Q.}\ \bibnamefont
  {Shen}}, \bibinfo {author} {\bibfnamefont {B.}~\bibnamefont {Gleason}},
  \bibinfo {author} {\bibfnamefont {G.~H.}\ \bibnamefont {McKinley}},\ and\
  \bibinfo {author} {\bibfnamefont {H.~A.}\ \bibnamefont {Stone}},\ }\bibfield
  {title} {\bibinfo {title} {Fiber coating with surfactant solutions},\
  }\href@noop {} {\bibfield  {journal} {\bibinfo  {journal} {Phys. Fluids}\
  }\textbf {\bibinfo {volume} {14}},\ \bibinfo {pages} {4055} (\bibinfo {year}
  {2002})}\BibitemShut {NoStop}%
\bibitem [{\citenamefont {Duprat}\ \emph {et~al.}(2007)\citenamefont {Duprat},
  \citenamefont {Ruyer-Quil}, \citenamefont {Kalliadasis},\ and\ \citenamefont
  {Giorgiutti-Dauphin{\'e}}}]{Duprat2007}%
  \BibitemOpen
  \bibfield  {author} {\bibinfo {author} {\bibfnamefont {C.}~\bibnamefont
  {Duprat}}, \bibinfo {author} {\bibfnamefont {C.}~\bibnamefont {Ruyer-Quil}},
  \bibinfo {author} {\bibfnamefont {S.}~\bibnamefont {Kalliadasis}},\ and\
  \bibinfo {author} {\bibfnamefont {F.}~\bibnamefont
  {Giorgiutti-Dauphin{\'e}}},\ }\bibfield  {title} {\bibinfo {title} {Absolute
  and convective instabilities of a viscous film flowing down a vertical
  fiber},\ }\href@noop {} {\bibfield  {journal} {\bibinfo  {journal} {Phys.
  Rev. Lett.}\ }\textbf {\bibinfo {volume} {98}},\ \bibinfo {pages} {244502}
  (\bibinfo {year} {2007})}\BibitemShut {NoStop}%
\bibitem [{\citenamefont {Blair}(1969)}]{Blair1969}%
  \BibitemOpen
  \bibfield  {author} {\bibinfo {author} {\bibfnamefont {G.~W.~S.}\
  \bibnamefont {Blair}},\ }\bibfield  {title} {\bibinfo {title} {Rheology and
  painting},\ }\href@noop {} {\bibfield  {journal} {\bibinfo  {journal}
  {Leonardo}\ ,\ \bibinfo {pages} {51}} (\bibinfo {year} {1969})}\BibitemShut
  {NoStop}%
\bibitem [{\citenamefont {Zenit}(2019)}]{Zenit2019}%
  \BibitemOpen
  \bibfield  {author} {\bibinfo {author} {\bibfnamefont {R.}~\bibnamefont
  {Zenit}},\ }\bibfield  {title} {\bibinfo {title} {Some fluid mechanical
  aspects of artistic painting},\ }\href@noop {} {\bibfield  {journal}
  {\bibinfo  {journal} {Phys. Rev. Fluids}\ }\textbf {\bibinfo {volume} {4}},\
  \bibinfo {pages} {110507} (\bibinfo {year} {2019})}\BibitemShut {NoStop}%
\bibitem [{\citenamefont {Chinju}\ \emph {et~al.}(2000)\citenamefont {Chinju},
  \citenamefont {Uchiyama},\ and\ \citenamefont {Mori}}]{Chinju2000}%
  \BibitemOpen
  \bibfield  {author} {\bibinfo {author} {\bibfnamefont {H.}~\bibnamefont
  {Chinju}}, \bibinfo {author} {\bibfnamefont {K.}~\bibnamefont {Uchiyama}},\
  and\ \bibinfo {author} {\bibfnamefont {Y.~H.}\ \bibnamefont {Mori}},\
  }\bibfield  {title} {\bibinfo {title} {“string-of-beads” flow of liquids
  on vertical wires for gas absorption},\ }\href@noop {} {\bibfield  {journal}
  {\bibinfo  {journal} {AIChE journal}\ }\textbf {\bibinfo {volume} {46}},\
  \bibinfo {pages} {937} (\bibinfo {year} {2000})}\BibitemShut {NoStop}%
\bibitem [{\citenamefont {Gr{\"u}nig}\ \emph {et~al.}(2012)\citenamefont
  {Gr{\"u}nig}, \citenamefont {Lyagin}, \citenamefont {Horn}, \citenamefont
  {Skale},\ and\ \citenamefont {Kraume}}]{Grunig2012}%
  \BibitemOpen
  \bibfield  {author} {\bibinfo {author} {\bibfnamefont {J.}~\bibnamefont
  {Gr{\"u}nig}}, \bibinfo {author} {\bibfnamefont {E.}~\bibnamefont {Lyagin}},
  \bibinfo {author} {\bibfnamefont {S.}~\bibnamefont {Horn}}, \bibinfo {author}
  {\bibfnamefont {T.}~\bibnamefont {Skale}},\ and\ \bibinfo {author}
  {\bibfnamefont {M.}~\bibnamefont {Kraume}},\ }\bibfield  {title} {\bibinfo
  {title} {Mass transfer characteristics of liquid films flowing down a
  vertical wire in a counter current gas flow},\ }\href@noop {} {\bibfield
  {journal} {\bibinfo  {journal} {Chem. Eng. Sci.}\ }\textbf {\bibinfo {volume}
  {69}},\ \bibinfo {pages} {329} (\bibinfo {year} {2012})}\BibitemShut
  {NoStop}%
\bibitem [{\citenamefont {Hosseini}\ \emph {et~al.}(2014)\citenamefont
  {Hosseini}, \citenamefont {Alizadeh}, \citenamefont {Fatehifar},\ and\
  \citenamefont {Alizadehdakhel}}]{Hosseini2014}%
  \BibitemOpen
  \bibfield  {author} {\bibinfo {author} {\bibfnamefont {S.~M.}\ \bibnamefont
  {Hosseini}}, \bibinfo {author} {\bibfnamefont {R.}~\bibnamefont {Alizadeh}},
  \bibinfo {author} {\bibfnamefont {E.}~\bibnamefont {Fatehifar}},\ and\
  \bibinfo {author} {\bibfnamefont {A.}~\bibnamefont {Alizadehdakhel}},\
  }\bibfield  {title} {\bibinfo {title} {Simulation of gas absorption into
  string-of-beads liquid flow with chemical reaction},\ }\href@noop {}
  {\bibfield  {journal} {\bibinfo  {journal} {Heat and Mass Transfer}\ }\textbf
  {\bibinfo {volume} {50}},\ \bibinfo {pages} {1393} (\bibinfo {year}
  {2014})}\BibitemShut {NoStop}%
\bibitem [{\citenamefont {Ding}\ \emph {et~al.}(2018)\citenamefont {Ding},
  \citenamefont {Xie}, \citenamefont {Ingham}, \citenamefont {Ma},\ and\
  \citenamefont {Pourkashanian}}]{Ding2018flow}%
  \BibitemOpen
  \bibfield  {author} {\bibinfo {author} {\bibfnamefont {H.}~\bibnamefont
  {Ding}}, \bibinfo {author} {\bibfnamefont {P.}~\bibnamefont {Xie}}, \bibinfo
  {author} {\bibfnamefont {D.}~\bibnamefont {Ingham}}, \bibinfo {author}
  {\bibfnamefont {L.}~\bibnamefont {Ma}},\ and\ \bibinfo {author}
  {\bibfnamefont {M.}~\bibnamefont {Pourkashanian}},\ }\bibfield  {title}
  {\bibinfo {title} {Flow behaviour of drop and jet modes of a laminar falling
  film on horizontal tubes},\ }\href@noop {} {\bibfield  {journal} {\bibinfo
  {journal} {Int. J. Heat Mass Transfer}\ }\textbf {\bibinfo {volume} {124}},\
  \bibinfo {pages} {929} (\bibinfo {year} {2018})}\BibitemShut {NoStop}%
\bibitem [{\citenamefont {Sadeghpour}\ \emph {et~al.}(2019)\citenamefont
  {Sadeghpour}, \citenamefont {Zeng}, \citenamefont {Ji}, \citenamefont
  {Dehdari~Ebrahimi}, \citenamefont {Bertozzi},\ and\ \citenamefont
  {Ju}}]{Sadeghpour2019}%
  \BibitemOpen
  \bibfield  {author} {\bibinfo {author} {\bibfnamefont {A.}~\bibnamefont
  {Sadeghpour}}, \bibinfo {author} {\bibfnamefont {Z.}~\bibnamefont {Zeng}},
  \bibinfo {author} {\bibfnamefont {H.}~\bibnamefont {Ji}}, \bibinfo {author}
  {\bibfnamefont {N.}~\bibnamefont {Dehdari~Ebrahimi}}, \bibinfo {author}
  {\bibfnamefont {A.~L.}\ \bibnamefont {Bertozzi}},\ and\ \bibinfo {author}
  {\bibfnamefont {Y.~S.}\ \bibnamefont {Ju}},\ }\bibfield  {title} {\bibinfo
  {title} {Water vapor capturing using an array of traveling liquid beads for
  desalination and water treatment},\ }\href@noop {} {\bibfield  {journal}
  {\bibinfo  {journal} {Science advances}\ }\textbf {\bibinfo {volume} {5}},\
  \bibinfo {pages} {eaav7662} (\bibinfo {year} {2019})}\BibitemShut {NoStop}%
\bibitem [{\citenamefont {Lee}\ \emph {et~al.}(2016)\citenamefont {Lee},
  \citenamefont {Brun}, \citenamefont {Marthelot}, \citenamefont {Balestra},
  \citenamefont {Gallaire},\ and\ \citenamefont {Reis}}]{Lee2016}%
  \BibitemOpen
  \bibfield  {author} {\bibinfo {author} {\bibfnamefont {A.}~\bibnamefont
  {Lee}}, \bibinfo {author} {\bibfnamefont {P.~T.}\ \bibnamefont {Brun}},
  \bibinfo {author} {\bibfnamefont {J.}~\bibnamefont {Marthelot}}, \bibinfo
  {author} {\bibfnamefont {G.}~\bibnamefont {Balestra}}, \bibinfo {author}
  {\bibfnamefont {F.}~\bibnamefont {Gallaire}},\ and\ \bibinfo {author}
  {\bibfnamefont {P.~M.}\ \bibnamefont {Reis}},\ }\bibfield  {title} {\bibinfo
  {title} {Fabrication of slender elastic shells by the coating of curved
  surfaces},\ }\href@noop {} {\bibfield  {journal} {\bibinfo  {journal} {Nature
  communications}\ }\textbf {\bibinfo {volume} {7}},\ \bibinfo {pages} {1}
  (\bibinfo {year} {2016})}\BibitemShut {NoStop}%
\bibitem [{\citenamefont {Zeng}\ \emph {et~al.}(2017)\citenamefont {Zeng},
  \citenamefont {Sadeghpour}, \citenamefont {Warrier},\ and\ \citenamefont
  {Ju}}]{Zeng2017}%
  \BibitemOpen
  \bibfield  {author} {\bibinfo {author} {\bibfnamefont {Z.}~\bibnamefont
  {Zeng}}, \bibinfo {author} {\bibfnamefont {A.}~\bibnamefont {Sadeghpour}},
  \bibinfo {author} {\bibfnamefont {G.}~\bibnamefont {Warrier}},\ and\ \bibinfo
  {author} {\bibfnamefont {Y.~S.}\ \bibnamefont {Ju}},\ }\bibfield  {title}
  {\bibinfo {title} {Experimental study of heat transfer between thin liquid
  films flowing down a vertical string in the rayleigh-plateau instability
  regime and a counterflowing gas stream},\ }\href@noop {} {\bibfield
  {journal} {\bibinfo  {journal} {Int. J. Heat Mass Transfer}\ }\textbf
  {\bibinfo {volume} {108}},\ \bibinfo {pages} {830} (\bibinfo {year}
  {2017})}\BibitemShut {NoStop}%
\bibitem [{\citenamefont {Zeng}\ \emph {et~al.}(2018)\citenamefont {Zeng},
  \citenamefont {Sadeghpour},\ and\ \citenamefont {Ju}}]{Zeng2018}%
  \BibitemOpen
  \bibfield  {author} {\bibinfo {author} {\bibfnamefont {Z.}~\bibnamefont
  {Zeng}}, \bibinfo {author} {\bibfnamefont {A.}~\bibnamefont {Sadeghpour}},\
  and\ \bibinfo {author} {\bibfnamefont {Y.~S.}\ \bibnamefont {Ju}},\
  }\bibfield  {title} {\bibinfo {title} {Thermohydraulic characteristics of a
  multi-string direct-contact heat exchanger},\ }\href@noop {} {\bibfield
  {journal} {\bibinfo  {journal} {Int. J. Heat Mass Transfer}\ }\textbf
  {\bibinfo {volume} {126}},\ \bibinfo {pages} {536} (\bibinfo {year}
  {2018})}\BibitemShut {NoStop}%
\bibitem [{\citenamefont {Eggers}\ and\ \citenamefont
  {Villermaux}(2008)}]{Eggers2008}%
  \BibitemOpen
  \bibfield  {author} {\bibinfo {author} {\bibfnamefont {J.}~\bibnamefont
  {Eggers}}\ and\ \bibinfo {author} {\bibfnamefont {E.}~\bibnamefont
  {Villermaux}},\ }\bibfield  {title} {\bibinfo {title} {Physics of liquid
  jets},\ }\href@noop {} {\bibfield  {journal} {\bibinfo  {journal} {Rep. Prog.
  Phys.}\ }\textbf {\bibinfo {volume} {71}},\ \bibinfo {pages} {036601}
  (\bibinfo {year} {2008})}\BibitemShut {NoStop}%
\bibitem [{\citenamefont {Gallaire}\ and\ \citenamefont
  {Brun}(2017)}]{Gallaire2017}%
  \BibitemOpen
  \bibfield  {author} {\bibinfo {author} {\bibfnamefont {F.}~\bibnamefont
  {Gallaire}}\ and\ \bibinfo {author} {\bibfnamefont {P.-T.}\ \bibnamefont
  {Brun}},\ }\bibfield  {title} {\bibinfo {title} {Fluid dynamic instabilities:
  theory and application to pattern forming in complex media},\ }\href@noop {}
  {\bibfield  {journal} {\bibinfo  {journal} {Philos. Trans. R. Soc. London,
  Ser. A}\ }\textbf {\bibinfo {volume} {375}},\ \bibinfo {pages} {20160155}
  (\bibinfo {year} {2017})}\BibitemShut {NoStop}%
\bibitem [{\citenamefont {Oron}\ \emph {et~al.}(1997)\citenamefont {Oron},
  \citenamefont {Davis},\ and\ \citenamefont {Bankoff}}]{Oron1997}%
  \BibitemOpen
  \bibfield  {author} {\bibinfo {author} {\bibfnamefont {A.}~\bibnamefont
  {Oron}}, \bibinfo {author} {\bibfnamefont {S.~H.}\ \bibnamefont {Davis}},\
  and\ \bibinfo {author} {\bibfnamefont {S.~G.}\ \bibnamefont {Bankoff}},\
  }\bibfield  {title} {\bibinfo {title} {Long-scale evolution of thin liquid
  films},\ }\href@noop {} {\bibfield  {journal} {\bibinfo  {journal} {Rev. Mod.
  Phys.}\ }\textbf {\bibinfo {volume} {69}},\ \bibinfo {pages} {931} (\bibinfo
  {year} {1997})}\BibitemShut {NoStop}%
\bibitem [{\citenamefont {Balestra}\ \emph {et~al.}(2019)\citenamefont
  {Balestra}, \citenamefont {Badaoui}, \citenamefont {Ducimeti\`{e}re},\ and\
  \citenamefont {Gallaire}}]{Balestra2019}%
  \BibitemOpen
  \bibfield  {author} {\bibinfo {author} {\bibfnamefont {G.}~\bibnamefont
  {Balestra}}, \bibinfo {author} {\bibfnamefont {M.}~\bibnamefont {Badaoui}},
  \bibinfo {author} {\bibfnamefont {Y.-M.}\ \bibnamefont {Ducimeti\`{e}re}},\
  and\ \bibinfo {author} {\bibfnamefont {F.}~\bibnamefont {Gallaire}},\
  }\bibfield  {title} {\bibinfo {title} {Fingering instability on curved
  substrates: optimal initial film and substrate perturbations},\ }\href@noop
  {} {\bibfield  {journal} {\bibinfo  {journal} {J. Fluid Mech.}\ }\textbf
  {\bibinfo {volume} {868}},\ \bibinfo {pages} {726} (\bibinfo {year}
  {2019})}\BibitemShut {NoStop}%
\bibitem [{\citenamefont {Hansen}\ and\ \citenamefont
  {Kelmanson}(1994)}]{Hansen1994}%
  \BibitemOpen
  \bibfield  {author} {\bibinfo {author} {\bibfnamefont {E.~B.}\ \bibnamefont
  {Hansen}}\ and\ \bibinfo {author} {\bibfnamefont {M.~A.}\ \bibnamefont
  {Kelmanson}},\ }\bibfield  {title} {\bibinfo {title} {Steady, viscous,
  free-surface flow on a rotating cylinder},\ }\href@noop {} {\bibfield
  {journal} {\bibinfo  {journal} {J. Fluid Mech.}\ }\textbf {\bibinfo {volume}
  {272}},\ \bibinfo {pages} {91} (\bibinfo {year} {1994})}\BibitemShut
  {NoStop}%
\bibitem [{\citenamefont {Peterson}\ \emph {et~al.}(2001)\citenamefont
  {Peterson}, \citenamefont {Jimack},\ and\ \citenamefont
  {Kelmanson}}]{Peterson2001}%
  \BibitemOpen
  \bibfield  {author} {\bibinfo {author} {\bibfnamefont {R.~C.}\ \bibnamefont
  {Peterson}}, \bibinfo {author} {\bibfnamefont {P.~K.}\ \bibnamefont
  {Jimack}},\ and\ \bibinfo {author} {\bibfnamefont {M.~A.}\ \bibnamefont
  {Kelmanson}},\ }\bibfield  {title} {\bibinfo {title} {On the stability of
  viscous free--surface flow supported by a rotating cylinder},\ }\href@noop {}
  {\bibfield  {journal} {\bibinfo  {journal} {Proceedings of the Royal Society
  of London. Series A: Mathematical, Physical and Engineering Sciences}\
  }\textbf {\bibinfo {volume} {457}},\ \bibinfo {pages} {1427} (\bibinfo {year}
  {2001})}\BibitemShut {NoStop}%
\bibitem [{\citenamefont {Ashmore}\ \emph {et~al.}(2003)\citenamefont
  {Ashmore}, \citenamefont {Hosoi},\ and\ \citenamefont {Stone}}]{Ashmore2003}%
  \BibitemOpen
  \bibfield  {author} {\bibinfo {author} {\bibfnamefont {J.}~\bibnamefont
  {Ashmore}}, \bibinfo {author} {\bibfnamefont {A.~E.}\ \bibnamefont {Hosoi}},\
  and\ \bibinfo {author} {\bibfnamefont {H.~A.}\ \bibnamefont {Stone}},\
  }\bibfield  {title} {\bibinfo {title} {The effect of surface tension on
  rimming flows in a partially filled rotating cylinder},\ }\href@noop {}
  {\bibfield  {journal} {\bibinfo  {journal} {J. Fluid Mech.}\ }\textbf
  {\bibinfo {volume} {479}},\ \bibinfo {pages} {65} (\bibinfo {year}
  {2003})}\BibitemShut {NoStop}%
\bibitem [{\citenamefont {Evans}\ \emph {et~al.}(2004)\citenamefont {Evans},
  \citenamefont {Schwartz},\ and\ \citenamefont {Roy}}]{Evans2004}%
  \BibitemOpen
  \bibfield  {author} {\bibinfo {author} {\bibfnamefont {P.~L.}\ \bibnamefont
  {Evans}}, \bibinfo {author} {\bibfnamefont {L.~W.}\ \bibnamefont
  {Schwartz}},\ and\ \bibinfo {author} {\bibfnamefont {R.~V.}\ \bibnamefont
  {Roy}},\ }\bibfield  {title} {\bibinfo {title} {Steady and unsteady solutions
  for coating flow on a rotating horizontal cylinder: Two-dimensional
  theoretical and numerical modeling},\ }\href@noop {} {\bibfield  {journal}
  {\bibinfo  {journal} {Phys. Fluids}\ }\textbf {\bibinfo {volume} {16}},\
  \bibinfo {pages} {2742} (\bibinfo {year} {2004})}\BibitemShut {NoStop}%
\bibitem [{\citenamefont {Li}\ and\ \citenamefont {Kumar}(2018)}]{Li2018}%
  \BibitemOpen
  \bibfield  {author} {\bibinfo {author} {\bibfnamefont {W.}~\bibnamefont
  {Li}}\ and\ \bibinfo {author} {\bibfnamefont {S.}~\bibnamefont {Kumar}},\
  }\bibfield  {title} {\bibinfo {title} {Three-dimensional surfactant-covered
  flows of thin liquid films on rotating cylinders},\ }\href@noop {} {\bibfield
   {journal} {\bibinfo  {journal} {J. Fluid Mech.}\ }\textbf {\bibinfo {volume}
  {844}},\ \bibinfo {pages} {61} (\bibinfo {year} {2018})}\BibitemShut
  {NoStop}%
\bibitem [{\citenamefont {Reisfeld}\ and\ \citenamefont
  {Bankoff}(1992)}]{Reisfeld1992}%
  \BibitemOpen
  \bibfield  {author} {\bibinfo {author} {\bibfnamefont {B.}~\bibnamefont
  {Reisfeld}}\ and\ \bibinfo {author} {\bibfnamefont {S.~G.}\ \bibnamefont
  {Bankoff}},\ }\bibfield  {title} {\bibinfo {title} {Non-isothermal flow of a
  liquid film on a horizontal cylinder},\ }\href@noop {} {\bibfield  {journal}
  {\bibinfo  {journal} {J. Fluid Mech.}\ }\textbf {\bibinfo {volume} {236}},\
  \bibinfo {pages} {167} (\bibinfo {year} {1992})}\BibitemShut {NoStop}%
\bibitem [{\citenamefont {Limat}\ \emph {et~al.}(1992)\citenamefont {Limat},
  \citenamefont {Jenffer}, \citenamefont {Dagens}, \citenamefont {Touron},
  \citenamefont {Fermigier},\ and\ \citenamefont {Wesfreid}}]{Limat1992}%
  \BibitemOpen
  \bibfield  {author} {\bibinfo {author} {\bibfnamefont {L.}~\bibnamefont
  {Limat}}, \bibinfo {author} {\bibfnamefont {P.}~\bibnamefont {Jenffer}},
  \bibinfo {author} {\bibfnamefont {B.}~\bibnamefont {Dagens}}, \bibinfo
  {author} {\bibfnamefont {E.}~\bibnamefont {Touron}}, \bibinfo {author}
  {\bibfnamefont {M.}~\bibnamefont {Fermigier}},\ and\ \bibinfo {author}
  {\bibfnamefont {J.~E.}\ \bibnamefont {Wesfreid}},\ }\bibfield  {title}
  {\bibinfo {title} {Gravitational instabilities of thin liquid layers:
  dynamics of pattern selection},\ }\href@noop {} {\bibfield  {journal}
  {\bibinfo  {journal} {Phys. D: Nonlinear Phenom.}\ }\textbf {\bibinfo
  {volume} {61}},\ \bibinfo {pages} {166} (\bibinfo {year} {1992})}\BibitemShut
  {NoStop}%
\bibitem [{\citenamefont {Weidner}(2013)}]{Weidner2013}%
  \BibitemOpen
  \bibfield  {author} {\bibinfo {author} {\bibfnamefont {D.~E.}\ \bibnamefont
  {Weidner}},\ }\bibfield  {title} {\bibinfo {title} {Suppression and reversal
  of drop formation on horizontal cylinders due to surfactant convection},\
  }\href@noop {} {\bibfield  {journal} {\bibinfo  {journal} {Phys. Fluids}\
  }\textbf {\bibinfo {volume} {25}},\ \bibinfo {pages} {082110} (\bibinfo
  {year} {2013})}\BibitemShut {NoStop}%
\bibitem [{\citenamefont {Rayleigh}(1883)}]{Rayleigh1882}%
  \BibitemOpen
  \bibfield  {author} {\bibinfo {author} {\bibfnamefont {L.}~\bibnamefont
  {Rayleigh}},\ }\bibfield  {title} {\bibinfo {title} {Investigation of the
  character of the equilibrium of an incompressible heavy fluid of variable
  density},\ }\href@noop {} {\bibfield  {journal} {\bibinfo  {journal} {Proc.
  London Math. Soc.}\ }\textbf {\bibinfo {volume} {s1-14}},\ \bibinfo {pages}
  {170} (\bibinfo {year} {1883})}\BibitemShut {NoStop}%
\bibitem [{\citenamefont {Taylor}(1950)}]{Taylor1950}%
  \BibitemOpen
  \bibfield  {author} {\bibinfo {author} {\bibfnamefont {G.~I.}\ \bibnamefont
  {Taylor}},\ }\bibfield  {title} {\bibinfo {title} {The instability of liquid
  surfaces when accelerated in a direction perpendicular to their planes. i},\
  }\href@noop {} {\bibfield  {journal} {\bibinfo  {journal} {Proc. R. Soc.
  London, Ser. A}\ }\textbf {\bibinfo {volume} {201}},\ \bibinfo {pages} {192}
  (\bibinfo {year} {1950})}\BibitemShut {NoStop}%
\bibitem [{\citenamefont {Plateau}(1873)}]{Plateau1873}%
  \BibitemOpen
  \bibfield  {author} {\bibinfo {author} {\bibfnamefont {J.~A.~F.}\
  \bibnamefont {Plateau}},\ }\href@noop {} {\emph {\bibinfo {title} {Statique
  exp{\'e}rimentale et th{\'e}orique des liquides soumis aux seules forces
  mol{\'e}culaires}}},\ Vol.~\bibinfo {volume} {2}\ (\bibinfo  {publisher}
  {Gauthier-Villars, Paris},\ \bibinfo {year} {1873})\BibitemShut {NoStop}%
\bibitem [{\citenamefont {Rayleigh}(1878)}]{Rayleigh1878}%
  \BibitemOpen
  \bibfield  {author} {\bibinfo {author} {\bibfnamefont {L.}~\bibnamefont
  {Rayleigh}},\ }\bibfield  {title} {\bibinfo {title} {On the instability of
  jets},\ }\href@noop {} {\bibfield  {journal} {\bibinfo  {journal} {Proc.
  London Math. Soc.}\ }\textbf {\bibinfo {volume} {1}},\ \bibinfo {pages} {4}
  (\bibinfo {year} {1878})}\BibitemShut {NoStop}%
\bibitem [{\citenamefont {Duclaux}\ \emph {et~al.}(2006)\citenamefont
  {Duclaux}, \citenamefont {Clanet},\ and\ \citenamefont
  {Qu{\'e}r{\'e}}}]{Duclaux2006}%
  \BibitemOpen
  \bibfield  {author} {\bibinfo {author} {\bibfnamefont {V.}~\bibnamefont
  {Duclaux}}, \bibinfo {author} {\bibfnamefont {C.}~\bibnamefont {Clanet}},\
  and\ \bibinfo {author} {\bibfnamefont {D.}~\bibnamefont {Qu{\'e}r{\'e}}},\
  }\bibfield  {title} {\bibinfo {title} {The effects of gravity on the
  capillary instability in tubes},\ }\href@noop {} {\bibfield  {journal}
  {\bibinfo  {journal} {J. Fluid Mech.}\ }\textbf {\bibinfo {volume} {556}},\
  \bibinfo {pages} {217} (\bibinfo {year} {2006})}\BibitemShut {NoStop}%
\bibitem [{\citenamefont {Eghbali}\ \emph {et~al.}(2023)\citenamefont
  {Eghbali}, \citenamefont {Ducimeti\`{e}re}, \citenamefont {Boujo},\ and\
  \citenamefont {Gallaire}}]{Eghbali2022in}%
  \BibitemOpen
  \bibfield  {author} {\bibinfo {author} {\bibfnamefont {S.}~\bibnamefont
  {Eghbali}}, \bibinfo {author} {\bibfnamefont {Y.~M.}\ \bibnamefont
  {Ducimeti\`{e}re}}, \bibinfo {author} {\bibfnamefont {E.}~\bibnamefont
  {Boujo}},\ and\ \bibinfo {author} {\bibfnamefont {F.}~\bibnamefont
  {Gallaire}},\ }\bibfield  {title} {\bibinfo {title} {Liquid film instability
  of an internally coated horizontal tube},\ }\href@noop {} {\bibfield
  {journal} {\bibinfo  {journal} {Phys. Rev. Fluids}\ }\textbf {\bibinfo
  {volume} {8}},\ \bibinfo {pages} {053901} (\bibinfo {year}
  {2023})}\BibitemShut {NoStop}%
\bibitem [{\citenamefont {Hooper}\ and\ \citenamefont
  {Boyd}(1983)}]{Hooper1983}%
  \BibitemOpen
  \bibfield  {author} {\bibinfo {author} {\bibfnamefont {A.~P.}\ \bibnamefont
  {Hooper}}\ and\ \bibinfo {author} {\bibfnamefont {W.~G.~C.}\ \bibnamefont
  {Boyd}},\ }\bibfield  {title} {\bibinfo {title} {Shear-flow instability at
  the interface between two viscous fluids},\ }\href@noop {} {\bibfield
  {journal} {\bibinfo  {journal} {J. Fluid Mech.}\ }\textbf {\bibinfo {volume}
  {128}},\ \bibinfo {pages} {507} (\bibinfo {year} {1983})}\BibitemShut
  {NoStop}%
\bibitem [{\citenamefont {Boomkamp}\ and\ \citenamefont
  {Miesen}(1996)}]{Boomkamp1996}%
  \BibitemOpen
  \bibfield  {author} {\bibinfo {author} {\bibfnamefont {P.~A.~M.}\
  \bibnamefont {Boomkamp}}\ and\ \bibinfo {author} {\bibfnamefont {R.~H.~M.}\
  \bibnamefont {Miesen}},\ }\bibfield  {title} {\bibinfo {title}
  {Classification of instabilities in parallel two-phase flow},\ }\href@noop {}
  {\bibfield  {journal} {\bibinfo  {journal} {International J. Multiphase
  Flow}\ }\textbf {\bibinfo {volume} {22}},\ \bibinfo {pages} {67} (\bibinfo
  {year} {1996})}\BibitemShut {NoStop}%
\bibitem [{\citenamefont {Kataoka}\ and\ \citenamefont
  {Troian}(1997)}]{Kataoka1997}%
  \BibitemOpen
  \bibfield  {author} {\bibinfo {author} {\bibfnamefont {D.~E.}\ \bibnamefont
  {Kataoka}}\ and\ \bibinfo {author} {\bibfnamefont {S.~M.}\ \bibnamefont
  {Troian}},\ }\bibfield  {title} {\bibinfo {title} {A theoretical study of
  instabilities at the advancing front of thermally driven coating films},\
  }\href@noop {} {\bibfield  {journal} {\bibinfo  {journal} {J. Colloid
  Interface Sci.}\ }\textbf {\bibinfo {volume} {192}},\ \bibinfo {pages} {350}
  (\bibinfo {year} {1997})}\BibitemShut {NoStop}%
\bibitem [{\citenamefont {Li}\ \emph {et~al.}(2011)\citenamefont {Li},
  \citenamefont {Yin},\ and\ \citenamefont {Yin}}]{Li2011}%
  \BibitemOpen
  \bibfield  {author} {\bibinfo {author} {\bibfnamefont {F.}~\bibnamefont
  {Li}}, \bibinfo {author} {\bibfnamefont {X.-Y.}\ \bibnamefont {Yin}},\ and\
  \bibinfo {author} {\bibfnamefont {X.-Z.}\ \bibnamefont {Yin}},\ }\bibfield
  {title} {\bibinfo {title} {Axisymmetric and non-axisymmetric instability of
  an electrically charged viscoelastic liquid jet},\ }\href@noop {} {\bibfield
  {journal} {\bibinfo  {journal} {J. non-newtonian Fluid Mech.}\ }\textbf
  {\bibinfo {volume} {166}},\ \bibinfo {pages} {1024} (\bibinfo {year}
  {2011})}\BibitemShut {NoStop}%
\bibitem [{\citenamefont {Fermigier}\ \emph {et~al.}(1992)\citenamefont
  {Fermigier}, \citenamefont {Limat}, \citenamefont {Wesfreid}, \citenamefont
  {Boudinet},\ and\ \citenamefont {Quilliet}}]{Fermigier1992}%
  \BibitemOpen
  \bibfield  {author} {\bibinfo {author} {\bibfnamefont {M.}~\bibnamefont
  {Fermigier}}, \bibinfo {author} {\bibfnamefont {L.}~\bibnamefont {Limat}},
  \bibinfo {author} {\bibfnamefont {J.~E.}\ \bibnamefont {Wesfreid}}, \bibinfo
  {author} {\bibfnamefont {P.}~\bibnamefont {Boudinet}},\ and\ \bibinfo
  {author} {\bibfnamefont {C.}~\bibnamefont {Quilliet}},\ }\bibfield  {title}
  {\bibinfo {title} {Two-dimensional patterns in rayleigh-taylor instability of
  a thin layer},\ }\href@noop {} {\bibfield  {journal} {\bibinfo  {journal} {J.
  Fluid Mech.}\ }\textbf {\bibinfo {volume} {236}},\ \bibinfo {pages} {349}
  (\bibinfo {year} {1992})}\BibitemShut {NoStop}%
\bibitem [{\citenamefont {Goren}(1962)}]{Goren1962}%
  \BibitemOpen
  \bibfield  {author} {\bibinfo {author} {\bibfnamefont {S.~L.}\ \bibnamefont
  {Goren}},\ }\bibfield  {title} {\bibinfo {title} {The instability of an
  annular thread of fluid},\ }\href@noop {} {\bibfield  {journal} {\bibinfo
  {journal} {J. Fluid Mech.}\ }\textbf {\bibinfo {volume} {12}},\ \bibinfo
  {pages} {309} (\bibinfo {year} {1962})}\BibitemShut {NoStop}%
\bibitem [{\citenamefont {Takagi}\ and\ \citenamefont
  {Huppert}(2010)}]{Takagi2010}%
  \BibitemOpen
  \bibfield  {author} {\bibinfo {author} {\bibfnamefont {D.}~\bibnamefont
  {Takagi}}\ and\ \bibinfo {author} {\bibfnamefont {H.~E.}\ \bibnamefont
  {Huppert}},\ }\bibfield  {title} {\bibinfo {title} {Flow and instability of
  thin films on a cylinder and sphere},\ }\href@noop {} {\bibfield  {journal}
  {\bibinfo  {journal} {J. Fluid Mech.}\ }\textbf {\bibinfo {volume} {647}},\
  \bibinfo {pages} {221} (\bibinfo {year} {2010})}\BibitemShut {NoStop}%
\bibitem [{\citenamefont {Carroll}(1984)}]{Carroll1984}%
  \BibitemOpen
  \bibfield  {author} {\bibinfo {author} {\bibfnamefont {B.~J.}\ \bibnamefont
  {Carroll}},\ }\bibfield  {title} {\bibinfo {title} {The equilibrium of liquid
  drops on smooth and rough circular cylinders},\ }\href@noop {} {\bibfield
  {journal} {\bibinfo  {journal} {J. Colloid Interface Sci.}\ }\textbf
  {\bibinfo {volume} {97}},\ \bibinfo {pages} {195} (\bibinfo {year}
  {1984})}\BibitemShut {NoStop}%
\bibitem [{\citenamefont {Brochard-Wyart}\ \emph {et~al.}(1990)\citenamefont
  {Brochard-Wyart}, \citenamefont {Di~Meglio},\ and\ \citenamefont
  {Qu{\'e}r{\'e}}}]{Brochard1990}%
  \BibitemOpen
  \bibfield  {author} {\bibinfo {author} {\bibfnamefont {F.}~\bibnamefont
  {Brochard-Wyart}}, \bibinfo {author} {\bibfnamefont {J.~M.}\ \bibnamefont
  {Di~Meglio}},\ and\ \bibinfo {author} {\bibfnamefont {D.}~\bibnamefont
  {Qu{\'e}r{\'e}}},\ }\bibfield  {title} {\bibinfo {title} {Theory of the
  dynamics of spreading of liquids on fibers},\ }\href@noop {} {\bibfield
  {journal} {\bibinfo  {journal} {Journal de Physique}\ }\textbf {\bibinfo
  {volume} {51}},\ \bibinfo {pages} {293} (\bibinfo {year} {1990})}\BibitemShut
  {NoStop}%
\bibitem [{\citenamefont {McHale}\ \emph {et~al.}(1999)\citenamefont {McHale},
  \citenamefont {Rowan}, \citenamefont {Newton},\ and\ \citenamefont
  {K\"ab}}]{McHale1999}%
  \BibitemOpen
  \bibfield  {author} {\bibinfo {author} {\bibfnamefont {G.}~\bibnamefont
  {McHale}}, \bibinfo {author} {\bibfnamefont {S.~M.}\ \bibnamefont {Rowan}},
  \bibinfo {author} {\bibfnamefont {M.~I.}\ \bibnamefont {Newton}},\ and\
  \bibinfo {author} {\bibfnamefont {N.~A.}\ \bibnamefont {K\"ab}},\ }\bibfield
  {title} {\bibinfo {title} {Estimation of contact angles on fibers},\
  }\href@noop {} {\bibfield  {journal} {\bibinfo  {journal} {J. adhesion
  science and technology}\ }\textbf {\bibinfo {volume} {13}},\ \bibinfo {pages}
  {1457} (\bibinfo {year} {1999})}\BibitemShut {NoStop}%
\bibitem [{\citenamefont {Aktershev}\ \emph {et~al.}(2021)\citenamefont
  {Aktershev}, \citenamefont {Alekseenko},\ and\ \citenamefont
  {Bobylev}}]{Aktershev2021}%
  \BibitemOpen
  \bibfield  {author} {\bibinfo {author} {\bibfnamefont {S.}~\bibnamefont
  {Aktershev}}, \bibinfo {author} {\bibfnamefont {S.}~\bibnamefont
  {Alekseenko}},\ and\ \bibinfo {author} {\bibfnamefont {A.}~\bibnamefont
  {Bobylev}},\ }\bibfield  {title} {\bibinfo {title} {Waves in a rivulet
  falling down an inclined cylinder},\ }\href@noop {} {\bibfield  {journal}
  {\bibinfo  {journal} {AIChE Journal}\ }\textbf {\bibinfo {volume} {67}},\
  \bibinfo {pages} {e17002} (\bibinfo {year} {2021})}\BibitemShut {NoStop}%
\bibitem [{\citenamefont {Craster}\ and\ \citenamefont
  {Matar}(2006)}]{Craster2006}%
  \BibitemOpen
  \bibfield  {author} {\bibinfo {author} {\bibfnamefont {R.~V.}\ \bibnamefont
  {Craster}}\ and\ \bibinfo {author} {\bibfnamefont {O.~K.}\ \bibnamefont
  {Matar}},\ }\bibfield  {title} {\bibinfo {title} {On viscous beads flowing
  down a vertical fibre},\ }\href@noop {} {\bibfield  {journal} {\bibinfo
  {journal} {J. Fluid Mech.}\ }\textbf {\bibinfo {volume} {553}},\ \bibinfo
  {pages} {85} (\bibinfo {year} {2006})}\BibitemShut {NoStop}%
\bibitem [{\citenamefont {Reynolds}(1886)}]{Reynolds1886}%
  \BibitemOpen
  \bibfield  {author} {\bibinfo {author} {\bibfnamefont {O.}~\bibnamefont
  {Reynolds}},\ }\bibfield  {title} {\bibinfo {title} {Iv. on the theory of
  lubrication and its application to mr. beauchamp tower’s experiments,
  including an experimental determination of the viscosity of olive oil},\
  }\href@noop {} {\bibfield  {journal} {\bibinfo  {journal} {Philos. Trans. R.
  Soc. London}\ }\textbf {\bibinfo {volume} {177}},\ \bibinfo {pages} {157}
  (\bibinfo {year} {1886})}\BibitemShut {NoStop}%
\bibitem [{\citenamefont {Eghbali}\ \emph {et~al.}(2022)\citenamefont
  {Eghbali}, \citenamefont {Keiser}, \citenamefont {Boujo},\ and\ \citenamefont
  {Gallaire}}]{Eghbali2022}%
  \BibitemOpen
  \bibfield  {author} {\bibinfo {author} {\bibfnamefont {S.}~\bibnamefont
  {Eghbali}}, \bibinfo {author} {\bibfnamefont {L.}~\bibnamefont {Keiser}},
  \bibinfo {author} {\bibfnamefont {E.}~\bibnamefont {Boujo}},\ and\ \bibinfo
  {author} {\bibfnamefont {F.}~\bibnamefont {Gallaire}},\ }\bibfield  {title}
  {\bibinfo {title} {Whirling instability of an eccentric coated fibre},\
  }\href@noop {} {\bibfield  {journal} {\bibinfo  {journal} {J. Fluid Mech.}\
  }\textbf {\bibinfo {volume} {952}},\ \bibinfo {pages} {A33} (\bibinfo {year}
  {2022})}\BibitemShut {NoStop}%
\end{thebibliography}%

\end{document}